\documentclass[preprint,review,12pt]{elsarticle}
\usepackage{enumerate}
\usepackage[usenames]{color}
\usepackage{bm}
\usepackage{graphicx}
\usepackage{amssymb} %The amssymb package provides various useful mathematical symbols
\usepackage{amsmath} % The amsthm package provides extended theorem environments
\journal{Nuclear Instruments and Methods in Physics
	}

\begin{document}

\begin{frontmatter}

\title{Neutral Kaon Spectrometer 2}

\author[aoba]	{M.~Kaneta\corref{cor}}
\ead{kaneta@lambda.phys.tohoku.ac.jp}
\author[aoba]	{B.~Beckford\fnref{umich}}
\author[aoba]	{T.~Fujii}
\author[aoba]	{Y.~Fujii\fnref{yakka}} 
\author[aoba]	{K.~Futatsukawa\fnref{kek_ac}}
\author[lan]	{Y.~C.~Han\fnref{inest}}
\author[aoba]	{O.~Hashimoto\fnref{deceased}} 
\author[aoba]	{K.~Hirose\fnref{jaea}} 
\author[elph]	{T.~Ishikawa}
\author[aoba]	{H.~Kanda\fnref{rcnp}}
\author[aoba]	{C.~Kimura} 
\author[aoba]	{K.~Maeda}
\author[aoba]	{S.~N.~Nakamura} 
\author[elph]	{K. Suzuki\fnref{wakasa}}
\author[aoba]	{K.~Tsukada\fnref{elph_p}}
\author[aoba]	{F.~Yamamoto} 
\author[elph]	{H.~Yamazaki\fnref{kek_rsc}}

\address[aoba]	{Department of Physics, Tohoku University, Sendai, Miyagi 980-8578, Japan}
\address[lan]	{School of Nuclear Science and Technology, Lanzhou University, Lanzhou 730000 China}
\address[elph]	{Research Center for Electron Photon Science (ELPH), Tohoku University, Sendai, Miyagi 982-0826, Japan}

\cortext[cor]{Corresponding author.}
\fntext[umich]	{Present Address: Department of Physics, University of Michigan, Ann Arbor, MI 48109-1040, USA}
\fntext[yakka]	{Present Address: Tohoku Medical and Pharmaceutical University, Sendai, Miyagi 981-8558, Japan}
\fntext[kek_ac]		{Present Address: Accelerator Laboratory, High Energy Accelerator Research Organization (KEK), Oho, Tsukuba, Ibaraki 305-0801, Japan}
\fntext[kek_rsc]		{Present Address: Radiation Science Center, High Energy Accelerator Research Organization (KEK), Tokai, Naka, Ibaraki 319-1195, Japan}
\fntext[inest]	{Present Address: Institute of Nuclear Energy Safety Technology (INEST), Chinese Academy of Science (CAS), Hefei, Anhui 230031, China} 
\fntext[deceased]	{deceased}
\fntext[jaea]		{Present Address: Advanced Science Research Center, Japan Atomic Energy Agency (JAEA), Tokai, Naka, Ibaraki 319-1195,  Japan}
\fntext[rcnp]		{Present address: Research Center for Nuclear Physics, Osaka University, Ibaraki, Osaka 567-0047, Japan}
\fntext[wakasa]	{Present Address: The Wakasa Wan Energy Research Center, Tsuruga, Fukui 914-0192, Japa}
\fntext[elph_p]	{Present Address: Research Center for Electron Photon Science (ELPH), Tohoku University, Sendai, Miyagi 982-0826,  Japan}

\begin{abstract}
  A large-acceptance spectrometer, Neutral Kaon Spectrometer 2 (NKS2), was newly constructed to explore various photoproduction reactions in the gigaelectronvolt region at the Laboratory of Nuclear Science (LNS, currently ELPH), Tohoku University.
  The spectrometer consisted of a dipole magnet, drift chambers, and plastic scintillation counters.
  NKS2 was designed to separate pions and protons in a momentum range of less than 1 GeV/$c$, and was placed in a tagged photon beamline.
  A cryogenic H$_{2}$/D$_{2}$ target fitted to the spectrometer were designed.
  The design and performance of the detectors are described.
  The results of the NKS2 experiment on analyzing strangeness photoproduction data using a 0.8--1.1 GeV tagged photon beam are also presented.

\end{abstract}

\begin{keyword}
  %\texttt{elsarticle.cls}\sep \LaTeX\sep Elsevier \sep template
  %\MSC[2010] 00-01\sep  99-00
  Magnetic spectrometer \sep
  Photonuclear reaction \sep
  Kaon \sep
  Pion \sep
  Tagged photon
\end{keyword}

\end{frontmatter}

%%%%%%%%%%%%%%%%%%
% \linenumbers

%%___________________________________________________________________________________________________________________

\section{Introduction}

  Photoproduction of the $K^{0}$ meson through the neutral channel, $\gamma+n \rightarrow K^0+\Lambda$, which has no charged particles in the reaction,
  provides key information on strangeness nuclear physics.
  Recent phenomenological models~\cite{David:1995pi, Mizutani:1997sd, Lee:1999kd} cannot explain photoproduction of $K^{+} + \Lambda$ from a proton
  and photoproduction of $K^{0} + \Lambda$ from a neutron consistently.
  For better understanding of strangeness photoproduction, high statistic and accurate experimental data, especially for neutral channels, are necessary.
  The first observation of photoproduction of neutral kaon events by a magnetic spectrometer was successfully carried out 
  using the TAGX spectrometer~\cite{Maruyama:1996aa, Maeda:2000aa} 
  at the 1.3 GeV electron synchrotron of the Institute for Nuclear Study, University of Tokyo.
  Neutral kaons were detected via the $K^{0}_{S} \rightarrow \pi^{+} \pi^{-}$ decay channel.
  However, the statistics of the report were significantly poor. 

  This lack of reliable experimental data for neutral kaon photoproduction encouraged the start of an experimental program 
  using the Neutral Kaon Spectrometer (NKS) at the Laboratory of Nuclear Science (LNS), Tohoku University,
  to study photoproduction of neutral kaons off deuterons.
  NKS was based on the TAGX spectrometer with modified counters and an augmented data acquisition (DAQ) system.

  The NKS collaboration performed pioneering experiments studying the $\gamma + d \rightarrow  K^{0} + \Lambda + X$~\cite{Tsukada:2008aa} 
  and $\gamma + ^{12}$C $\rightarrow K^{0} + \Lambda + X$~\cite{Watanabe2007photo} reactions near the threshold energy for strangeness photoproduction.
  The threshold energy $E_{\gamma}$ is 0.914 GeV for the elementary $\gamma + n \rightarrow K^0 + \Lambda$ reaction assuming the free neutron target at rest.
  The NKS collaboration reported that the cross-section of $K^0$ photoproduction showed similar behavior as previously reported $K^+$ data.

  Though the NKS spectrometer played an important role in the pioneering works, it had some issues to be addressed.
  The spectrometer had no detectors on the beamline to reduce the material thickness, and thus it had less sensitivity in the forward region.
  Based on the experience of the NKS experiments, 
  the second-generation neutral kaon spectrometer (NKS2) was newly constructed to have a larger acceptance and better spatial resolution.
  Established technologies were fully utilized for construction of NKS2.

  The following are the requirements of the NKS2 experiment:
  The relative momentum resolutions ($\sigma_{p}/p$) for pions and protons are better than 5~\% in 1 GeV/$c$ region to describe differential cross-section as a function of momentum.
  The invariant mass width is better than 5 MeV/$c^{2}$ for the $K^{0}$ and $\Lambda$ measurements.
  We also require clean $\pi/p$ separation as a part of our particle identification capabilities to reconstruct 
  $K^{0}_{S}$ from $\pi^{+} \pi^{-}$ decay and $\Lambda$ from $p \pi^{-}$ decay.
  The decay vertex resolution should be good enough to determine the target region.

  Strangeness photoproduction data were taken with a liquid deuterium target from 2006 to 2007.
  The number of tagged photons on the target was $4.0 \times 10^{12}$ in that experimental run period.
  The data analysis was already completed, and NKS2 showed the expected performance.
  The results of the experiment were reported in Refs.~\cite{Futatsukawa:2011aa, Futatsukawa:2012aa}

  We upgraded the inner detectors from 2007 to 2008 and commissioned them in 2009.
  We gathered data in 2010 using liquid hydrogen and deuterium targets. 
  The recorded numbers of tagged photons on the target were $3.1 \times 10^{11}$ and $8.9 \times 10^{11}$ for the liquid hydrogen and deuterium targets, respectively.
  From the analysis performed on those data sets, we reported the $\Lambda$ photoproduction measurements in the $\gamma$+$d$ reaction in Refs.~\cite{Beckford:2012aa, Beckford:2016aa}.
  In the following sections, we describe the structural details and performance of NKS2.

  \begin{figure*}[tbh]
	\begin{center}
		\includegraphics[bb=0 0 284 208,width=10cm,clip]{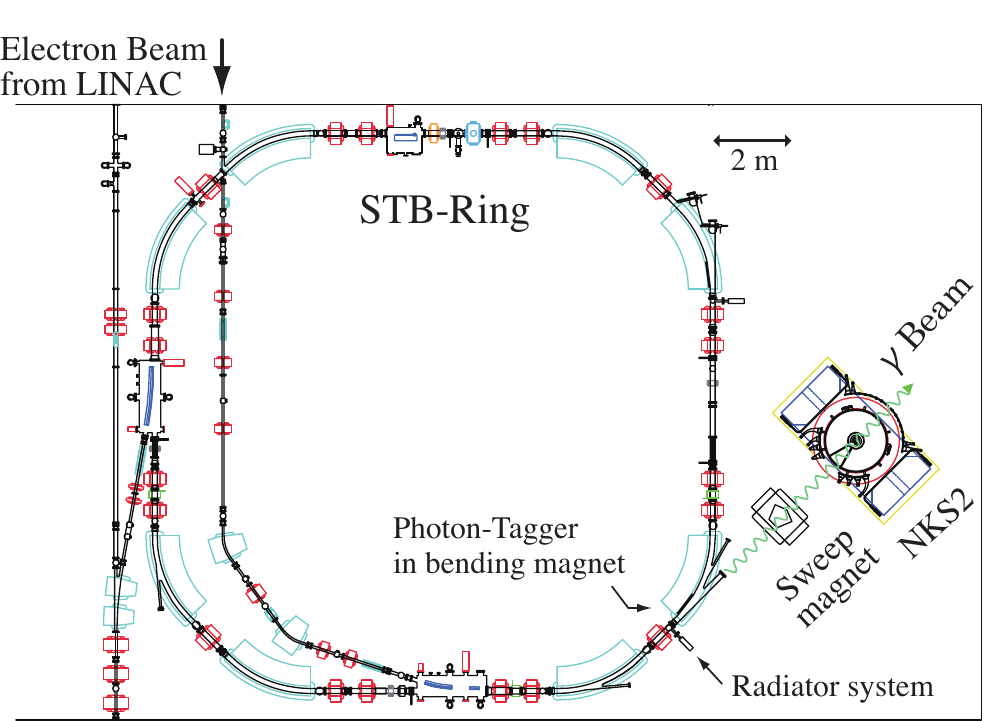}
		\caption{Floor plan in the second experimental hall at LNS-Tohoku.
			 The electron synchrotron STB-ring, a photon-tagger system, a sweep magnet, and NKS2 are placed.}
		\label{fig:floor_plan}
	\end{center}
  \end{figure*}

%%___________________________________________________________________________________________________________________

\section{Specification of NKS2}

  NKS was replaced by NKS2 at LNS-Tohoku\footnote{Present Name: Research Center of Electron Photon Science, Tohoku University (ELPH)} in 2004.
  NKS2 was placed in the second experimental hall at LNS-Tohoku (see Fig.~\ref{fig:floor_plan}).
  NKS2 consisted of a dipole magnet, two types of drift chambers: vertex drift chamber (VDC) and cylindrical drift chamber (CDC), 
  two sets of plastic scintillation hodoscopes: inner hodoscope (IH) and outer hodoscope (OH), and electron veto (EV) counters.

\begin{figure*}[htb]
	\begin{center}
		\includegraphics[bb=0 0 455 249,width=10cm,clip]{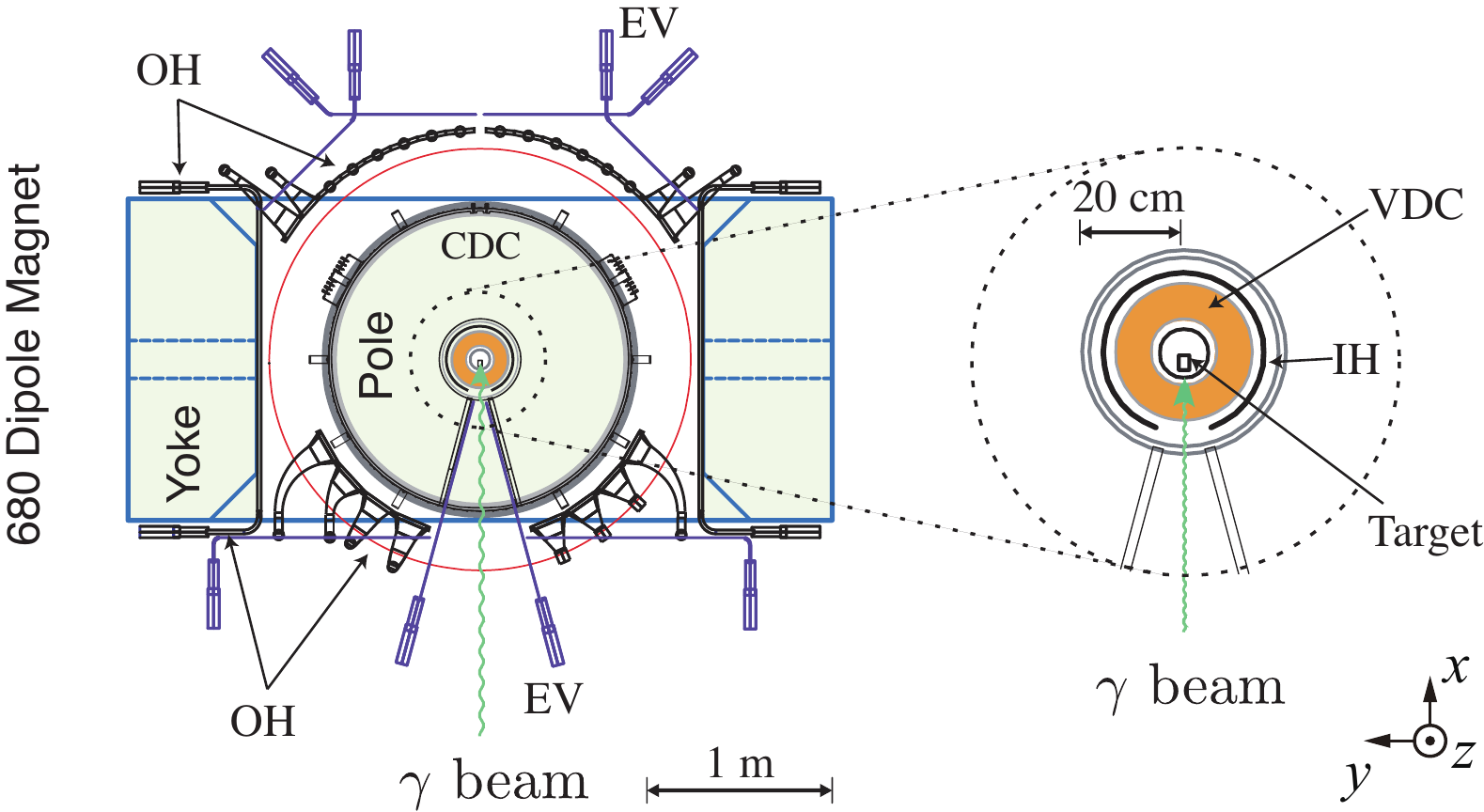}
		\caption{
			Schematic view of NKS2.
			Inner detectors are shown in the zoomed image.
			NKS2 consisted of the 680 dipole magnet,  the drift chambers (VDC and CDC), the hodoscopes (IH and OH), the EV counters, and the target.
			}
		\label{fig:nks2}
	\end{center}
  \end{figure*}

  Figure~\ref{fig:nks2} shows the schematic view of the detector alignment. 
  The target was surrounded by these NKS2 detectors.
  The detectors were inserted between the poles of a dipole magnet with an aperture of 680~mm. 
  Descriptions of the magnet, the drift chambers, the hodoscopes, the EV counters, and the target are shown in Sections~\ref{sec:magnet}, ~\ref{sec:chambers},  ~\ref{sec:hodoscopes}, ~\ref{sec:ev}, and ~\ref{sec:target}, respectively.

%____________________________________________

\subsection{680 Dipole Magnet}
  \label{sec:magnet}

\begin{figure}[htbp]
	\begin{center}
		\includegraphics[bb=0 0 1370 2491,width=7.5cm,clip]{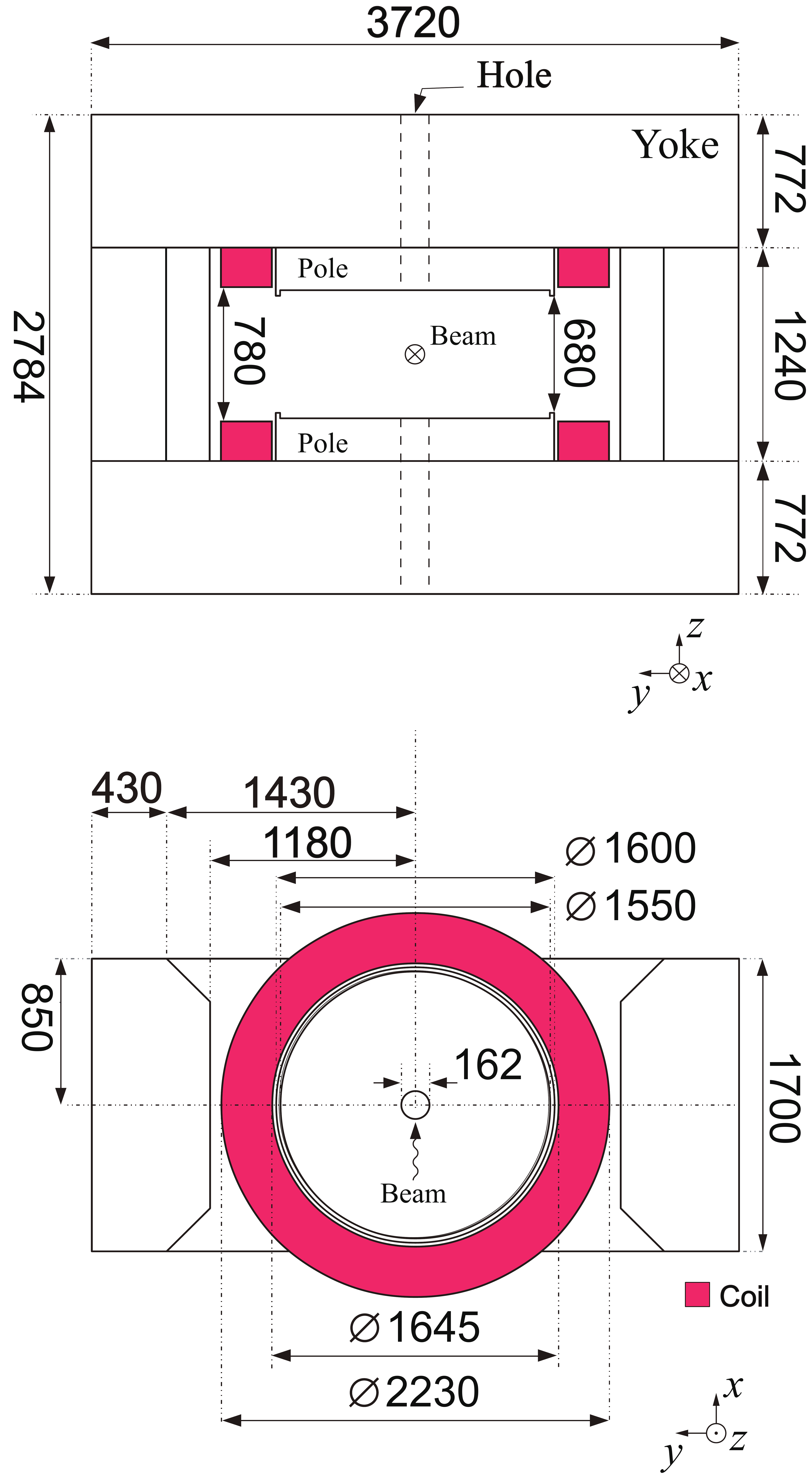}
		\caption{
			Schematic view of the dipole magnet. 
			The top (bottom) figure shows the cross-section along the vertical (horizontal) direction.
			The unit is millimeter.
			The coil was made of stack of five pieces (flat ``pancake" coil) at each pole.
			}
		\label{fig:680_magnet}
	\end{center}
  \end{figure}

  The 680 magnet was named by its gap size of 680 mm.
  The schematic view of the 680 magnet is shown in Fig.~\ref{fig:680_magnet}.
  Table~\ref{table:680_magnet} is a summary of the specifications and operation conditions of the magnet.

  The 680 magnet was previously used as a cyclotron magnet at the Cyclotron Radioisotope Center (CYRIC), Tohoku University.
  Additional iron pieces were added to extend its pole gap.
  The height of the gap was determined by optimization of the maximum acceptance 
  and a sufficiently strong magnetic field to achieve the required momentum resolution.

\begin{figure}[htbp]
	\begin{center}
		\includegraphics[bb=0 0 220 220,width=7.5cm,clip]{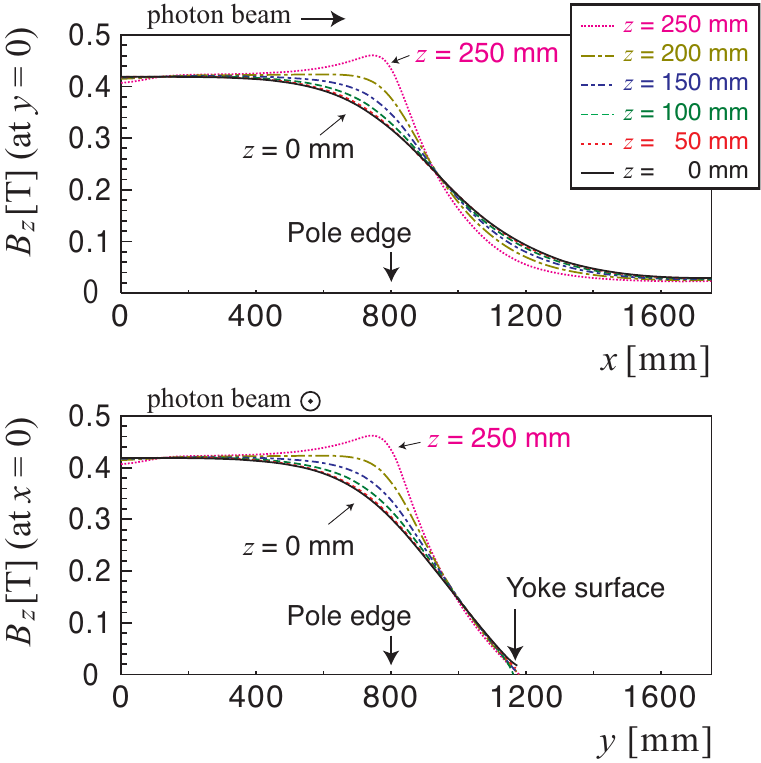}
		\caption{
			Magnetic field distribution calculated with TOSCA.
			The top figure shows the field strength along the beam direction ($x$). 
			The bottom figure shows the strength along the direction perpendicular to the beamline ($y$).
			The lines in both figures correspond to different $z$ positions
			($z$ is the opposite direction of gravity and the origin of the coordinate is the center of the magnet).
			The definition of the coordinate is also shown in Fig.~\ref{fig:680_magnet}.
			}
		\label{fig:magnetic_field}
	\end{center}
  \end{figure}

  We used a 3D finite-element-method calculation (OPERA-3d/TOSCA~\cite{Simkin:1979aa, Simkin:1980aa}) to obtain a magnetic field map.
  The 3D magnetic field map with a 2 cm mesh size was constructed from the TOSCA calculation result in order to track charged particles in NKS2.
  The results are shown in Fig.~\ref{fig:magnetic_field}.
  The strength of the magnetic field was 0.42~T at the center.
  We measured the field strength along the beamline to calibrate the calculated magnetic field. 

  We checked the stability of the magnetic field at the center of the magnet with a Hall probe (Panasonic Semiconductor OH10008).
  After ten minutes from excitation of the magnet, the relative field fluctuation was less than $10^{-5}$.

  The dipole magnet had two holes of 162 mm {\o} at the center of the top and bottom of the yoke.
  The top-side hole was used for the target insertion.
  We set a web camera to monitor the target cell position from the bottom-side hole (see Section~\ref{sec:target}).

  \begin{table}[htbp]
	\begin{center}
		\caption{
			Specification and operation condition of the 680 dipole magnet.
			}
		\label{table:680_magnet}
		\begin{tabular}{l l}
			\hline
			Pole gap		&	680 mm \\
			Pole radius		&	800 mm \\
			Number of coil units	&	10 \\
			Number of turns in a coil unit		&	25  \\
			Yoke size			&	$3720^{\mathrm{W}}$ mm $\times 1700^{\mathrm{D}}$ mm $\times 2784^{\mathrm{H}}$ mm \\
			Total weight		&	110 t \\
			Operation current		&	1000 A \\
			Magnetic field at the center	&	0.42 T \\
			\hline
		\end{tabular}
	\end{center}
  \end{table}

%_______________________________________

\subsection{Drift Chambers}
  \label{sec:chambers}

  The track of a charged particle was determined with the combined information from a pair of drift chambers, 
  the Vertex Drift Chamber (VDC) and the Cylindrical Drift Chamber (CDC).
  They were placed in the gap of the 680 magnet. 
  The premixed gas of argon (50~\%) and ethane (50~\%) was used at the atmospheric pressure.
  The sense wires were made of gold plated tungsten-rhenium with $20~\mu$m {\o}.
  The shield and field wires are $100~\mu$m  {\o} and made of gold-plated copper-beryllium.

%____

\subsubsection{VDC}

\begin{figure}[htbp]
	\begin{center}
		\includegraphics[bb=0 0 1554 3038,width=7.5cm, clip]{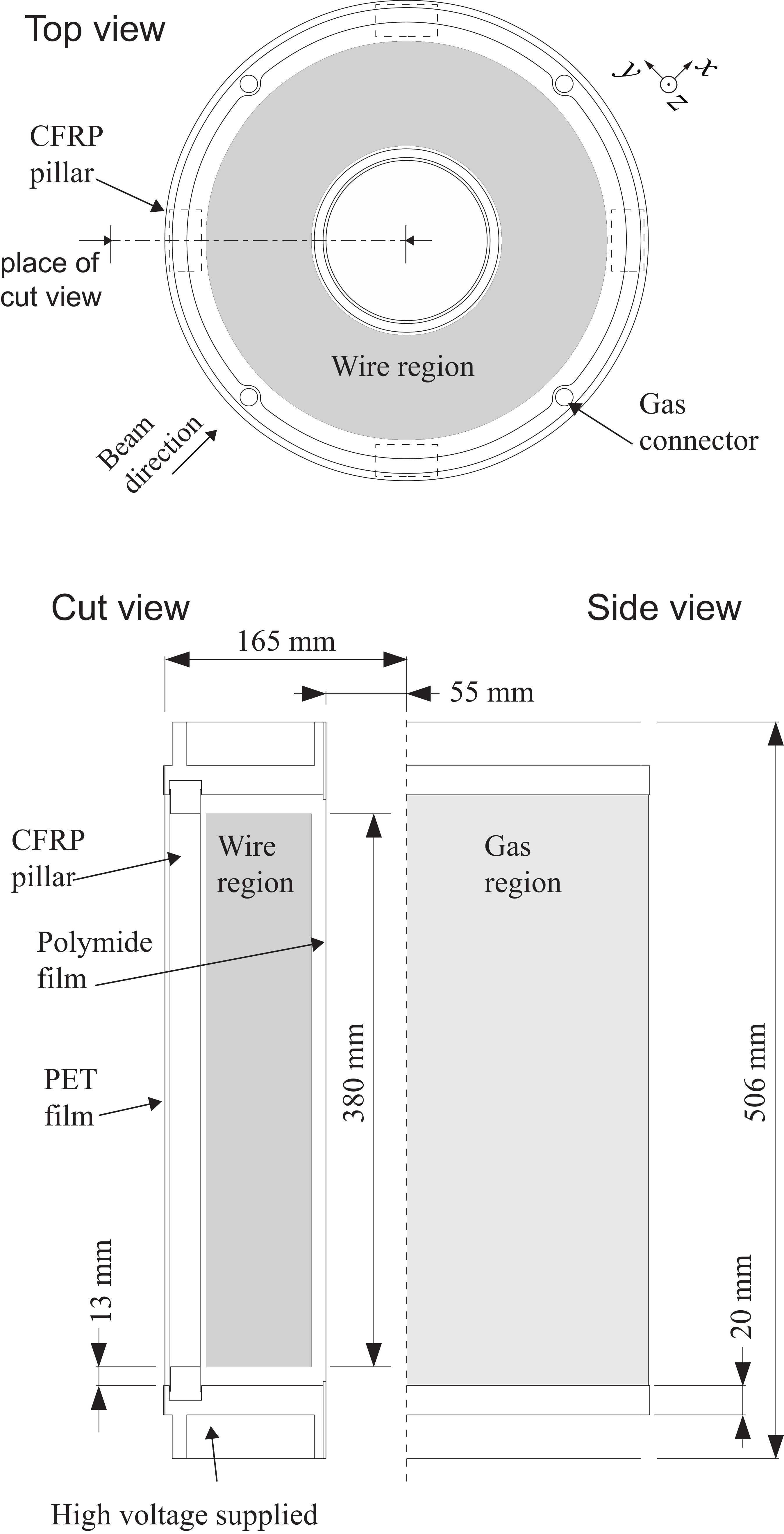}
		\caption{
			Schematic view of VDC.
	        	The cross-section in the vertical direction (indicated by the dotted-dashed line in the top view) is shown in the bottom left.
	        	}
		\label{fig:VDC}
	\end{center}
  \end{figure}

  Figure~\ref{fig:VDC} shows the schematic view of VDC.
  The specifications and operation conditions are summarized in Table~\ref{table:VDC}.
  The outer diameter of VDC was 330~mm with a height of 506~mm.
  The detector was composed of 626 sense wires placed at stereo angles such that they created eight layers (u, u', v, v', u, u', v, v'). 
  The stereo angle was defined as the angle with respect to the gravity direction.

  \begin{table*}[htbp]
	\begin{center}
		\caption{Specification and operation condition of VDC}
		\label{table:VDC}
		\begin{tabular}{l l}
			\hline
			Drift gas		&	Ar+C$_{2}$H$_{6}$ (50:50) \\
			Gas pressure	&	1 atm \\
			Body material 	&	Aluminum alloy (JIS A5052) \\
			Pillar material		&	0.5-mm-thick CFRP  \\
			Body size			&	110--330 mm {\o} $\times 506^{\mathrm{H}}$ mm \\
			Number of layers		&	8 (all stereo layers) \\
			Stereo angle of sense wire	&	6.18--10.83$^{\circ}$ \\
			Sense wire			&	20 $\mu$m {\o} Au plated W (Re doped) \\
			Field and shield wire	&	100  $\mu$m {\o} Au plated Cu--Be \\
			Wire tension			&	0.39 N (sense) \\
								&	0.78 N (field and shield) \\
			Cell shape			&	Trapezoid \\
			Half-cell size			&	4 mm \\
			Operation voltage		&	$-2.1$ kV (field) and  $-1.4$ kV (shield) \\
			\hline
		\end{tabular}
	\end{center}
  \end{table*}

  The cells were trapezoidal in shape with a half-cell size of approximately 4 mm (4 mm pitch in the radial direction at the end-plate).
  The wires were fixed with a 3~mm~{\o} feedthrough~\cite{Sekimoto:2004sy} by soldering.
  The tension of the wires was 0.39~N for the sense wire and 0.78~N for the field and shield wires.
  The wire tension was respectively loaded with a weight of 40 g and 80 g during the construction.
  The body was supported by the four pillars made of carbon-fiber-reinforced-plastic (CFRP) in order to withstand the tension of the wires.

\begin{figure}[htbp]
	\begin{center}
		\includegraphics[bb=0 0 517 551,width=7.5cm, clip]{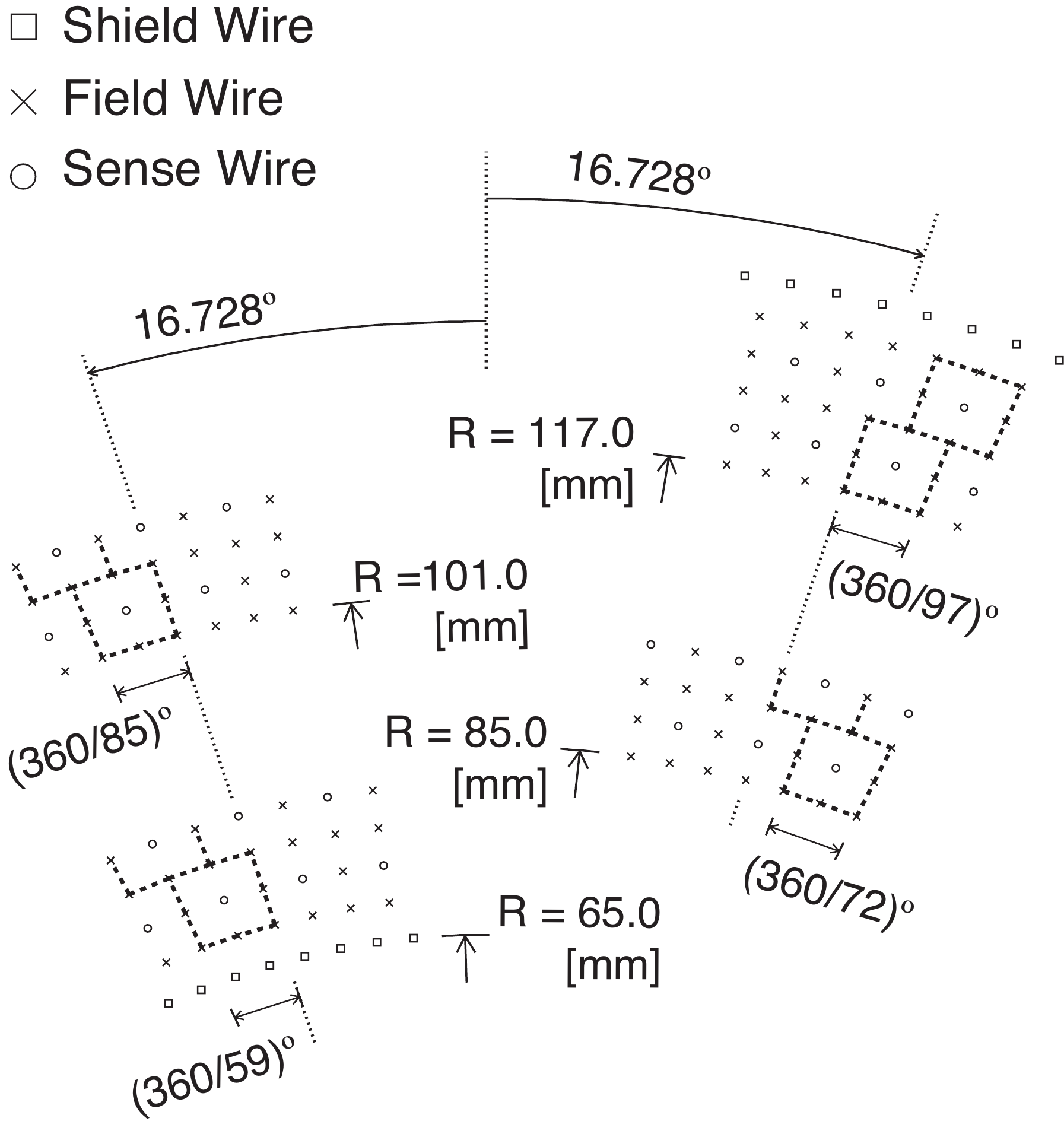}
		\caption{
			Cell structure of VDC. 
			A part of cells corresponding to the beamline is shown.
			The wires on the dotted lines were crossed with the $x$ axis (beamline) at $z=0$. 
	 		The half-cell size was 4~mm in the radial direction.
			R refers to the distance of the feedthrough position from the center on the end-plate. 
			The shield wires were shifted by a quarter-cell from the field wires in the angular direction.
			The angle coverage of one cell is shown for each layer group and the denominator corresponds to the number of cells.
			}
	\label{fig:VDC_cell_structure}
	\end{center}
  \end{figure}

\begin{figure}[htbp]
	\begin{center}
		\includegraphics[bb=0 0 198 184,width=7.5cm, clip]{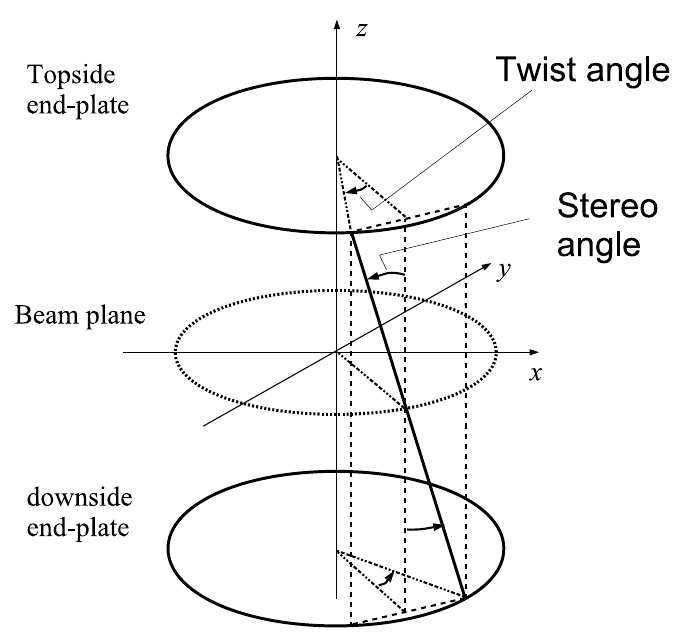}
		\caption{
			Definition of the twist and the stereo angles.
			The feedthrough were placed along the circles on the upside and downside end-plates.
			}
	\label{fig:twist_stereo_definition}
	\end{center}
  \end{figure}

  The cell structure of VDC is shown in Fig.~\ref{fig:VDC_cell_structure}.
  The position of the wires corresponds to the holes for the feedthrough at the top side of the end-plate.
  A twist angle of 16.728 degrees was the common value for all layers.
  The value was chosen to make the stereo angle of the inner shield wire to be 6.0 degrees.
  The definitions of the twist angle and the stereo angle are illustrated in Fig.~\ref{fig:twist_stereo_definition}.
  The u, u' (v, v') layers were twisted counterclockwise (clockwise). 
  The stereo angle of the sense wires increased from 6.18 to 10.83 degrees with the radius R (distance from VDC center to the wire).

  VDC had no shield wires between the u, u' layer group and the v, v' layer group to increase the sensitive region (see Fig.~\ref{fig:VDC_cell_structure}) by reducing the number of wires except the most inner and most outer layers.
  The outside field wires of u' (v') cells are given by the field wires of the next v (u) cell except for the most outer v' cells.

  The operation voltage was $-2.1$ kV for the field wires.
  Based on the result of drift chamber simulation code GARFIELD~\cite{Veenhof:1993hz}, we supplied 2/3 of the field wire voltage of the shield wires.
  A high-voltage (HV) scan using cosmic rays was performed from $-1.6$ to $-2.3$ kV for the field wires to determine the operating voltage. 
  Two plastic scintillation counters were used to select cosmic rays.
  The HV dependence of the singles rate showed a plateau behavior around $-2.0$ to $-2.1$ kV, and the layer efficiency was more than 98~\% in the region.
  We chose $-2.1$ kV to minimize gain drift caused by effective voltage drops in a high-count rate environment.

  VDC was originally designed to have symmetric wire alignment; 
  hence, there was a significant amount of wire materials on the beamline upstream of the target. 
  They contributed to photon conversion and caused the background.
  In order to reduce the background rate, the wires in the upstream direction along the beamline were removed.

\subsubsection{CDC}
 
  CDC was designed with a diameter of 1600 mm and a height of 630 mm 
  and consisted of ten layers (x, x', u, u', x, x', v, v', x, x').
  The axial wires were parallel to the gravity direction and were denoted by x and x'.
  The stereo wires were tilted, and the stereo angles were defined with respect to the gravity direction (denoted by u, u', v, and v').

\begin{figure}[htbp]
	\begin{center}
		\includegraphics[bb=0 0 562 895,width=7.5cm, clip]{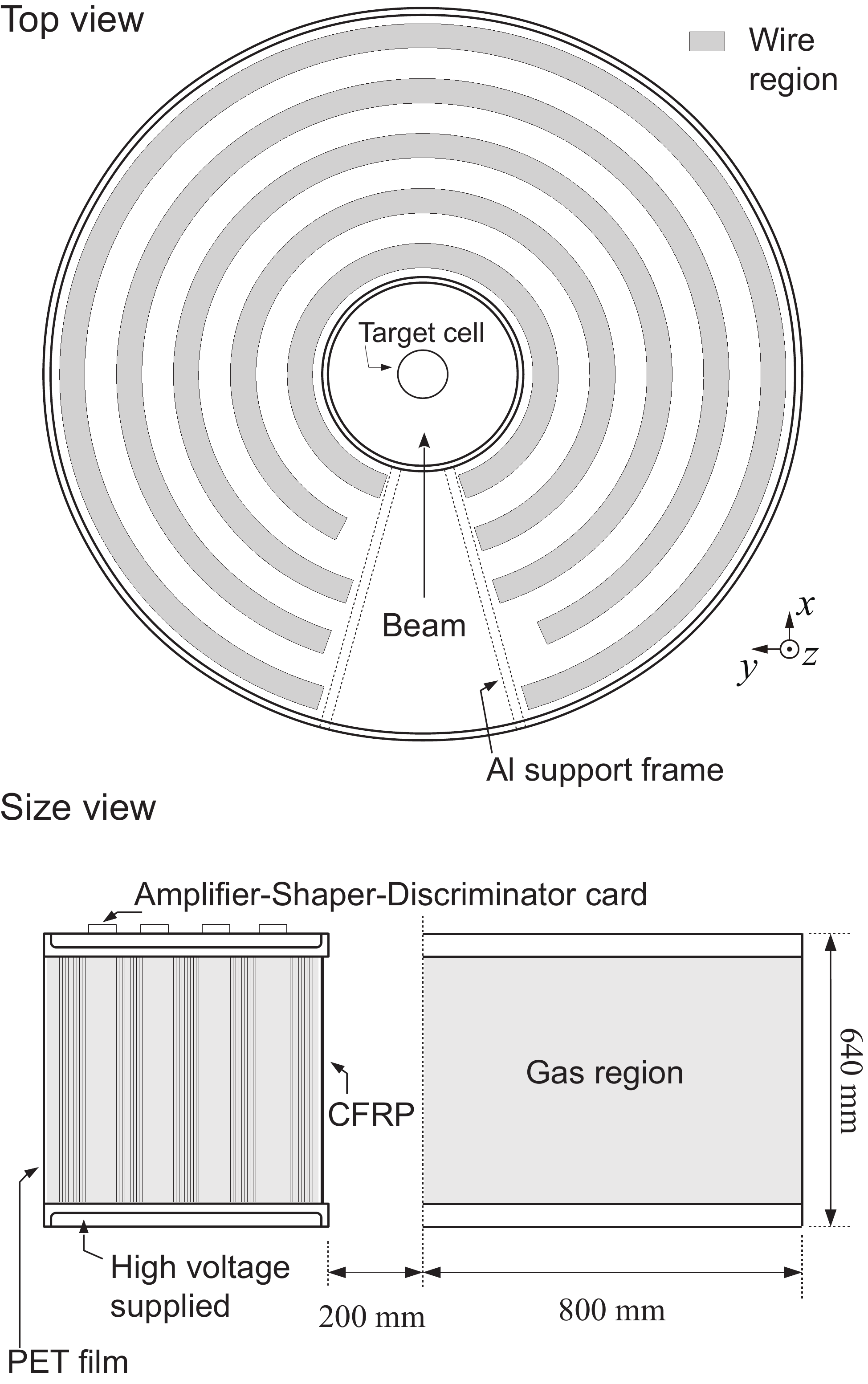}
		\caption{
			 Schematic view of CDC.
		         The cross-section in the vertical direction along the beam is shown in the bottom left.
			}
		\label{fig:CDC}
	\end{center}
  \end{figure}

  Figure~\ref{fig:CDC} shows the body structure of CDC.
  The aluminum end-plate of CDC was supported by a CFRP cylinder and an aluminum alloy support frame.
  For easy installation of CDC in the gap of the 680 magnet, CDC was placed on rails coated by 2-mm-thick Teflon, and it was manually moved.

\begin{figure}[htbp]
	\begin{center}
		\includegraphics[bb=0 0 312 726,width=7.5cm, clip]{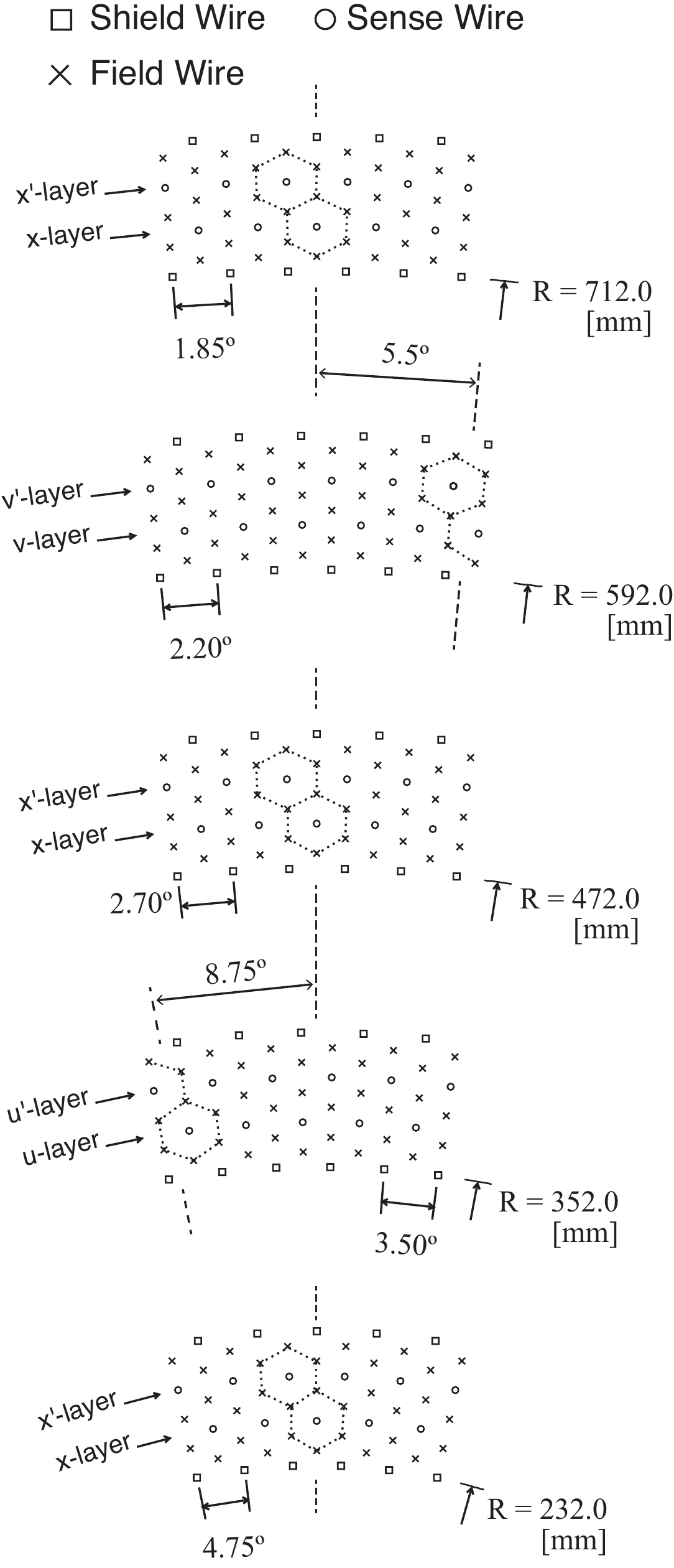}
		\caption{
			Cell structure of CDC.
			The markers show the feedthrough position on the end-plate.
			The dashed lines correspond to the beamline at $z$ = 0.
			The half-cell size is 12 mm in the radial direction.
			R refers to the distance between the feedthrough position and the center on the end-plate. 
	        	}
		\label{fig:CDC_cell_structure}
	\end{center}
  \end{figure}

  The ten layers were further defined into groups consisting of two layers within each group.
  The u, u' (v, v') stereo layers were twisted counterclockwise (clockwise) by 8.75 (5.50) degrees toward the $z$ axis (the gravity direction).
  The stereo angles of u, u', v, and v' were 6.44, 6.75, 6.68, and 6.87 degrees, respectively.
  The cell shapes are shown in Fig.~\ref{fig:CDC_cell_structure}.
 
  The angular coverage of the chamber on the beam plane was approximately from $-162$ to 162 degrees by axial wires and from $-158$ to 158 degrees by stereo wires
  (the angle is counted from the beam direction).
  The radial coverage was from 238 to 760 mm over all layers. 
  The cell had a hexagonal honeycomb structure, and the maximum drift length was approximately 12 mm.
 
  Under the operating condition during the experiment, 
  an HV of $-2.8$~kV was supplied to the field wires and $-1.4$~kV to the shield wires (half of the field wire voltage).
  The voltage of the shield wires was chosen from the results of the GARFIELD study to have the rotational-symmetrical field shape around the sense wire.
 
  The wires were fixed by a 4~mm~{\o} feedthrough with soldering.
  The sense wires were tensioned with 0.49~N (loaded with a weight of 50 g), and the shield and field wires were tensioned with 0.78~N (loaded with a weight of 80 g).
 
  The combination of VDC and CDC allowed the tracking of charged particles in three dimensions.
  They were essential to measure the momentum and determine the decay vertex points of $K^{0}_{S}$ and $\Lambda$.

  \begin{table*}[tbhp]
	\begin{center}
		\caption{Specification and operation condition of CDC}
		\label{table:CDC}
		\begin{tabular}{l l}
			\hline
			Drift gas		&	Ar+C$_{2}$H$_{6}$ (50:50) \\
			Gas pressure	&	1 atm \\
			Body material 	&	Aluminum alloy (JIS A5052) \\
			Support cylinder material	&	0.5-mm-thick CFRP  \\
			Body size			&	400--1600 mm {\o} $\times 630^{\mathrm{H}}$ mm \\
			Number of layers		&	10 (6 axial and 4 stereo layers) \\
			Stereo angle			&	6.44--6.87$^{\circ}$ \\
			Sense wire			&	20 $\mu$m {\o} Au plated W (Re doped) \\
			Field and shield wire	&	100  $\mu$m {\o} Au plated Cu-Be \\
			Wire tension			&	0.49 N (sense) \\
								&	0.78 N (field and shield) \\
			Cell shape			&	hexagon\\
			Half-cell size			&	12 mm \\
			Operation voltage		&	$-2.8$ kV (field) and  $-1.4$ kV (shield) \\
			\hline
		\end{tabular}
	\end{center}
  \end{table*}

\subsubsection{Readout Electronics}
  \label{sec:DC_readout}

  Amplifier-shaper-discriminator (ASD) cards equipped with a SONY CXA3183Q chip~\cite{ATLAS_ASD_chip} were used for chamber readout.
  The gain was 0.8 V/pC at preamp, and there were two models with different integration time constants (16 ns and 80 ns).
  We chose 16 ns integration time constant types, because a high hit rate was expected.
  The ASD chip had analog and digital (LVDS) outputs, and we used the digital signal.

  The timing information of the chambers was recorded by AMT-VME~\cite{Arai:2001tw} modules with the common stop mode.
  The module had a time-to-digital converter (TDC) LSI chip, which was developed for the ATLAS Muon TDC (AMT) module~\cite{Arai:2000eb, Arai:2001ka}, and the least time count was 0.78 ns.

  We developed the firmware of the module to set a timing window via the VME memory access.
  The module recorded hits even after a common stop signal with the original firmware, and the recorded event size became large.
  The date acquisition (DAQ) efficiency was about 10~\% at a trigger rate of 2 kHz before the firmware modification.
  After optimization of the time window with the new firmware, it became approximately 70~\% at a trigger rate of 2 kHz.

%__________________________________________

\subsection{Timing Counters}
  \label{sec:hodoscopes}

  A charged particle was identified by its momentum and velocity.
  The velocity was computed from the time of flight (TOF) and the flight length. 
  We had two sets of hodoscopes as TOF counters: IH and OH. 
  Each hodoscope consisted of a solid plastic scintillator and two photomultiplier tubes (PMTs).
  IH and OH also were used as multiplicity counters in the trigger level.
 
\subsubsection{IH}

  IH was added into the logic as a start counter of the TOF measurement and served as the time reference in the event trigger.
  Each segment was attached to a light guide and a fine mesh dynode type PMT, HAMAMATSU Photonics H6152-01B at each end.
  It operated with a negative HV setting.
  The fine-mesh dynode type PMT allows operation within a magnetic field of 0.42 T supplied by the 680 magnet.

  \begin{figure}[thbp]
    \begin{center}
      \includegraphics[bb=0 0 150 192,width=7.5cm, clip]{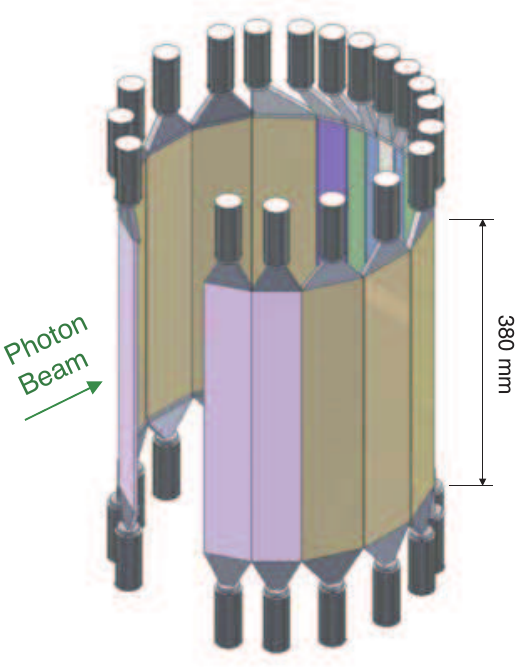}
      \caption{
			3D schematic view of IH.
			IH is mounted on VDC (a supporter is not shown).
			The cross-section of IH on the beam plane is shown in Fig.~\ref{fig:IH_scinti}.
	}
      \label{fig:IH}
    \end{center}
  \end{figure}

\begin{figure}[htbp]
    \begin{center}
      \includegraphics[bb=0 0 427 481,width=5.5cm, clip]{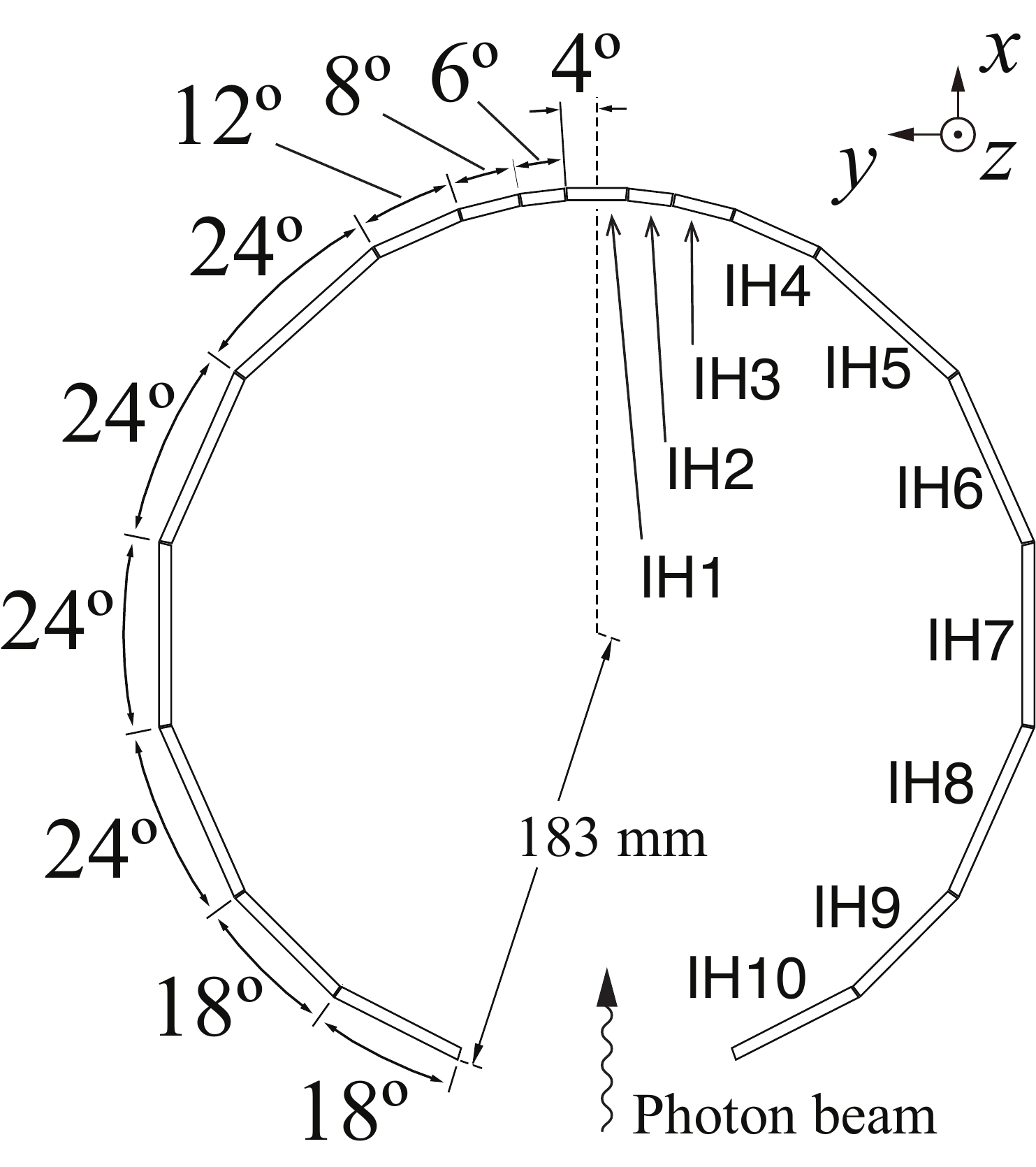}
      \caption{
			Figure shows the positions of the plastic scintillators of IH.
			IH counters were categorized into left and right arms (left/right was defined as seeing from upstream of the beamline) except for IH1.
			The edge of each plastic scintillator was located on a circle with a radius of 183 mm.
			}
      \label{fig:IH_scinti}
    \end{center}
  \end{figure}

  Figures~\ref{fig:IH} and \ref{fig:IH_scinti} show the arrangement of IH counters. Table~\ref{table:IH} is a summary of IH specification.
  IH counter segments were arranged to enclose VDC.
  The length of the scintillator bars of IH was similar to the height of VDC sensitive area to cover the vertical acceptance within a geometric condition.
  We adopted a double-side readout for IH counters (IH2--IH10 at the left and right side), except for IH1.
  There was a gap of $\pm$25 mm between upside and downside of IH1 to avoid photon beam hits that will generate $e^{+} e^{-}$ background.
  The scintillator thickness was 5 mm for all of the counters.

  The scintillator bars were wrapped by a 25-$\mu$m thick aluminized Mylar film.
  The light guide and a part of the PMT were surrounded by a Teflon tape.
  The counter was finally wrapped by a black tape for light shielding.

  The size of each segment was determined considering the probability of multiple hits and the singles rate on one segment.
  The design requirement was a probability of less than 2~\% of multihits and the singles rate $\leq$ 200 kHz.
  From the result of a GEANT4~\cite{Geant4:2003aa, Geant4:2006aa, Geant4:2016aa} simulation, 
  incoming bremsstrahlung photons of an energy from 5 MeV to 1.2 GeV gave a singles rate of 160 kHz at a tagged photon rate of 2 MHz.
  Under the real beam condition, the maximum singles rate on each counter was less than 200 kHz as we expected.

  \begin{table}[htbp]
	\begin{center}
		\caption{Specification of IH}
		\label{table:IH}
		\begin{tabular}{l l}
			\hline
			PMT					&	HAMAMATSU Photonics, H6152-01B \\
			Plastic scintillator			&	ELJEN Technology, EJ-230 \\
			Scintillator geometry			&	\\
			\multicolumn{1}{r}{cross-section shape} & \\
			\multicolumn{1}{r}{in the beam plane}	&	trapezoid \\
			\multicolumn{1}{r}{thickness}		&	5 mm \\
			\multicolumn{1}{r}{height}		&	165 mm for IH1 \\
								&	380 mm for IH2 to IH10 \\
			\hline
		\end{tabular}
	\end{center}
  \end{table}

\subsubsection{OH}
 
\begin{figure*}[htbp]
    \begin{center}
      \includegraphics[bb=0 0 508 736,width=11cm]{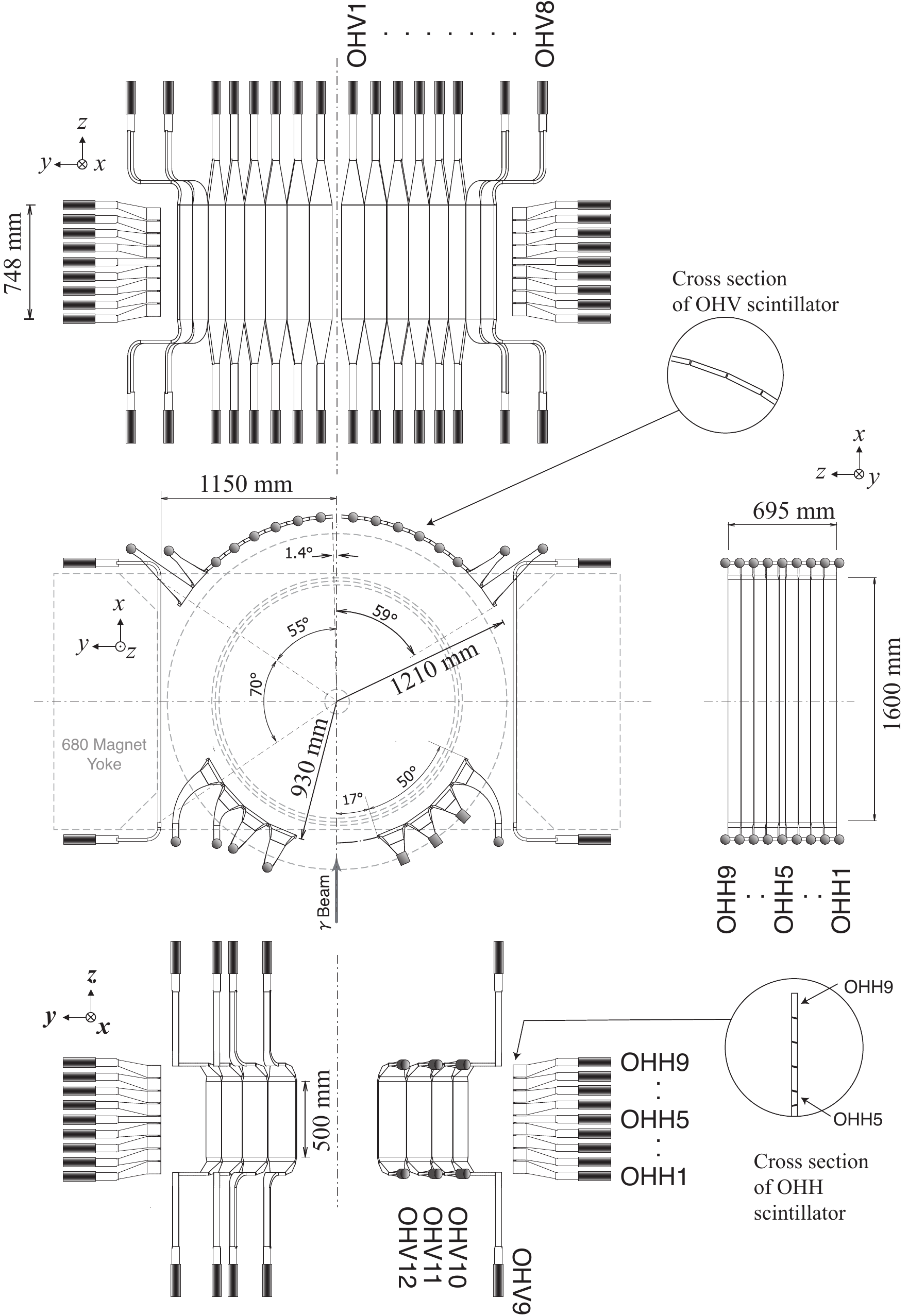}
      \caption{
        Schematic view of OH.
        OH counters were categorized into left and right arms (left/right was defined as seeing from upstream of the beamline).
        OHV was arranged vertically downstream of the target (OHV1--OHV8) and in upstream (OHV9--OHV12).
        OHHs were installed in the position of the symmetry of the $y-z$ plane. 
      }
      \label{fig:OH}
    \end{center}
  \end{figure*}

  OH was used as a stop counter of the TOF.
  OH consisted of two sections.
  One section was arranged vertically (OHV) and the other comprised horizontal bars (OHH).
  Those were categorized into the left and right components. 
  The orientation of OH counters is shown in Fig.~\ref{fig:OH}. 
  The specification of OH is listed in Table~\ref{table:OH}.
  
  OHV consisted of 24 segments in total, and they were categorized into two groups separated by the beamline.
  They were OH vertical left (OHVL) and right (OHVR).
  OHV1--OHV8 (OHV9--OHV12) used a plastic scintillator~\footnote{The manufacturer and model number were not recognized} 
  of $748^{\mathrm{H}} \times 150^{\mathrm{W}}\times 20^{\mathrm{T}}$ ($500^{\mathrm{H}} \times 200^{\mathrm{W}} \times 20^{\mathrm{T}}$)~mm$^{3}$.

  OHH had nine segments each on the left and right side.
  They were arranged in parallel to the beam direction.
  OHH1 and OHH9 (OHH2--OHH4 and OHH6--OHH8) used a plastic scintillator 
  of $1600^{\mathrm{L}} \times 82.5^{\mathrm{H}} \times 20^{\mathrm{T}}$ ($1600^{\mathrm{L}} \times 80^{\mathrm{H}} \times 20^{\mathrm{T}}$)~mm$^{3}$.
  The height of OHH5 scintillator bar was reduced to 45~mm in order to keep the count rate at an acceptable level.

  We designed the light guide to set the PMT on the surface of the magnet yoke except for OHVR10--OHVR12.
  Because there were the power supply and cooling water lines for the magnet, it was required for OHVR10--OHVR12 to set the PMT in the magnetic field.
  We used HAMAMATSU Photonics R5924-70 (2-inch fine mesh dynode type PMT) for those counters.
  For the other OHs, 2 inch PMTs (HAMAMATSU Photonics, H1161 or H7195) were used.
  The method of light shielding of OH was the same as for IH.

  \begin{table*}[htbp]
	\begin{center}
		\caption{Specification of OH}
		\label{table:OH}
		\begin{tabular}{l l}
			\hline
			PMT											&	HAMAMATSU Photonics,\\
															&	R5924-70 for OHVR10--12 and \\
															&	H1161 and H7195 for the others \\
			Scintillator geometry								& \\
			\multicolumn{1}{r}{cross-section shape}				&	trapezoid (OHV and OHH5) \\
															&	parallelogram (OHH1--4, 6--9) \\
			\multicolumn{1}{r}{thickness}						&	20 mm \\
			\multicolumn{1}{r}{height and long edge of trapezoid}	&	748 mm and 150 mm (OHV1--8) \\
															&	500 mm and 200 mm (OHV9--12) \\
			\multicolumn{1}{r}{length and long edge of trapezoid}	&	1600 mm and 45 mm for (OHH5) \\
			\multicolumn{1}{r}{length and edge of parallelogram}	&	1600 mm and 80 mm for (OHH2--4, 6--8) \\
															&	1600 mm and 82.5 mm for (OHH1, 9) \\
			\hline
		\end{tabular}
	\end{center}
  \end{table*}

\subsection{EV counters} 
  \label{sec:ev}

  The EV counters were originally designed to reject $e^{+}e^{-}$ pairs produced by photon conversion. 
  They were placed on the beam plane.
  One EV unit consisted of a plastic scintillator and one PMT because we had no requirement of high timing resolution for EV counters.
  The EV counters were placed upstream and downstream of the 680 magnet (see Fig.~\ref{fig:nks2}). 

  The downstream EV counter was placed outside of the forward OHV counters and consisted of four plastic scintillation counters of $1125^{\mathrm{L}} \times 50^{\mathrm{W}} \times 10^{\mathrm{T}}$~mm$^3$.
  The upstream EV counter was placed outside of the backward OHV counters and consisted of two plastic scintillation counters of $1250^{\mathrm{L}} \times 50^{\mathrm{W}} \times 5^{\mathrm{T}}$~mm$^3$.
  The other set was placed in the opening of the backward region of CDC and consisted of two plastic scintillators of $600^{\mathrm{L}} \times 150^{\mathrm{W}} \times 10^{\mathrm{T}}$~mm$^3$.

  The upstream EV counter was placed to reject the $e^{+}e^{-}$ background that was produced upstream of the spectrometer and was not rejected by the sweep magnet.
  The downstream EV counter originally had the role to reject the background produced at the target.
  We planned to use both sides of the EV counters in a trigger level to reduce the trigger rate and to increase the DAQ rate.
  However, the NKS collaboration reported that it made a bias in kinematics~\cite{Tsukada:2005aa}.
  They used downstream EV counter in the trigger, 
  because the NKS DAQ system could not satisfy a sufficient efficiency under 2--3 kHz trigger.

  After commissioning NKS2, the NKS2 DAQ system performed with an efficiency of 70~\%
  under a 2 kHz trigger rate without a downstream EV counter.
  We decided to take the data without a downstream EV counter, but with the upstream EV counter in the trigger level.
  The set of downstream EV counters was used as reference counters to record the singles rate on the beam plane. 

%%___________________________________________________________________________________________________________________

\section{Cryogenic Target System} 
  \label{sec:target}

  A cryogenic target system for hydrogen or deuterium was installed in NKS2.
  The target cryostat was a cylindrical vacuum chamber composed of two parts:
  a thick part (300~{\o} $\times 300^{\mathrm{L}}$~mm) and a thin part (100~{\o} $\times 1500^{\mathrm{L}}$~mm).

  The thick part held a two-staged Gifford--McMahon refrigerator Sumitomo Heavy Industries (SHI) RD-208B inside.
  A turbo-molecular pump (TMP), Leibold Turbovac 151 was directly connected to the vacuum chamber to keep the pressure $\sim 10^{-5}$ Pa during operation of the target.

\begin{figure*}[tbhp]
    \begin{center}
      \includegraphics[bb=0 0 435 295,width=8cm,clip]{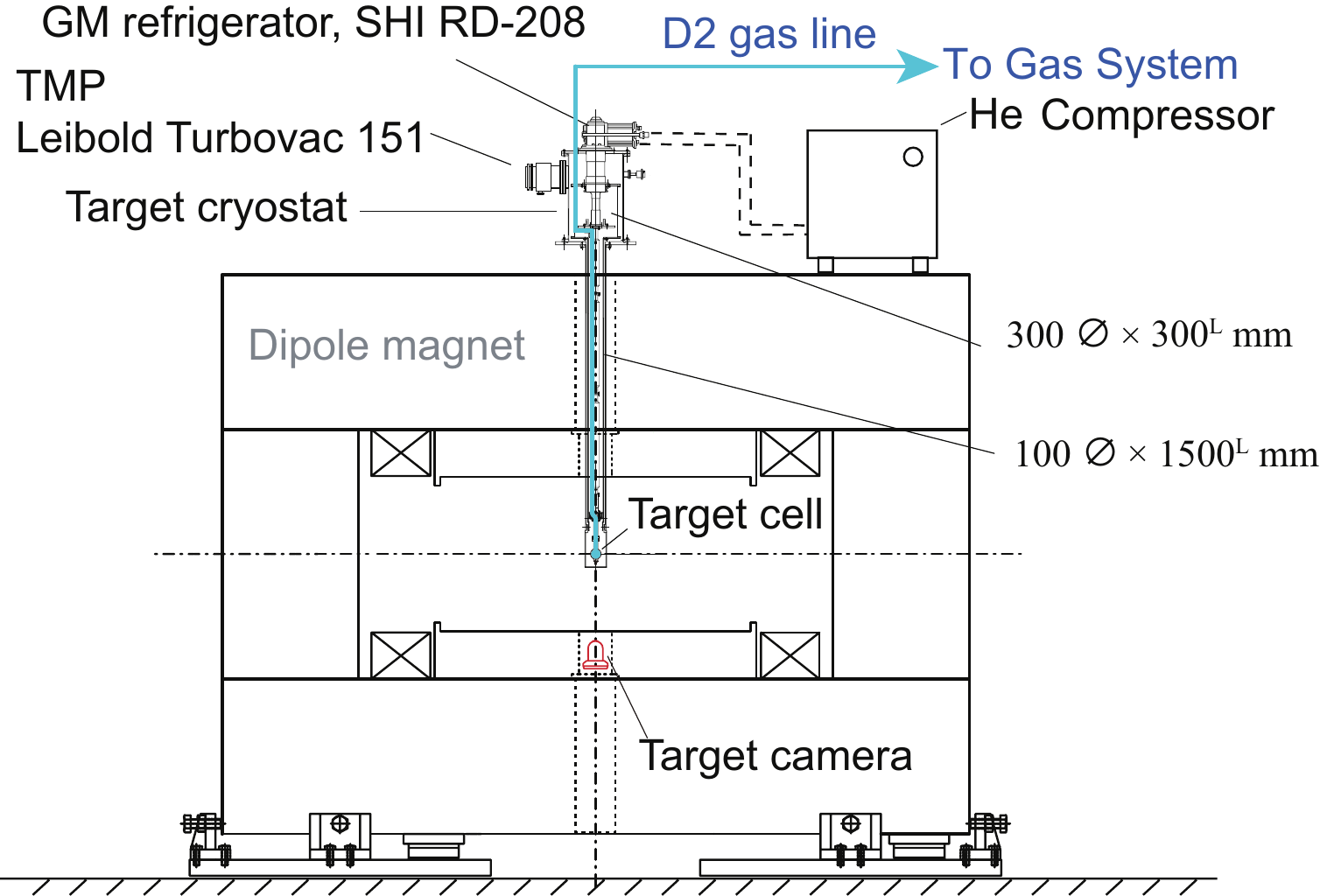}
      \caption{
		Schematic drawing of the dipole magnet and the target system.
		The target cryostat was inserted in the vertical hole of the dipole magnet.
		The target camera viewed the target position from the bottom of the cryostat.
		}
      \label{fig:cryostat}
    \end{center}
  \end{figure*}

  The thin part was shaped for insertion into the vertical hole of the dipole magnet and held a target cell.
  Liquid D$_{2}$/H$_{2}$ was stored in the target cell placed on the bottom of the cryostat.
  The target position was aligned to the center of a pole gap of the dipole magnet.
  The schematic view of the system is shown in Fig.~\ref{fig:cryostat}.

  The position of the cryostat was viewed by a web-camera from the bottom side.
  It monitored the target position and was utilized to check the relative position between the target cryostat
  and the inner wall of VDC when the vacuum chamber was moved in.

\begin{figure}[htbp]
    \begin{center}
      \includegraphics[bb=0 0 770 451,width=6.5cm,clip]{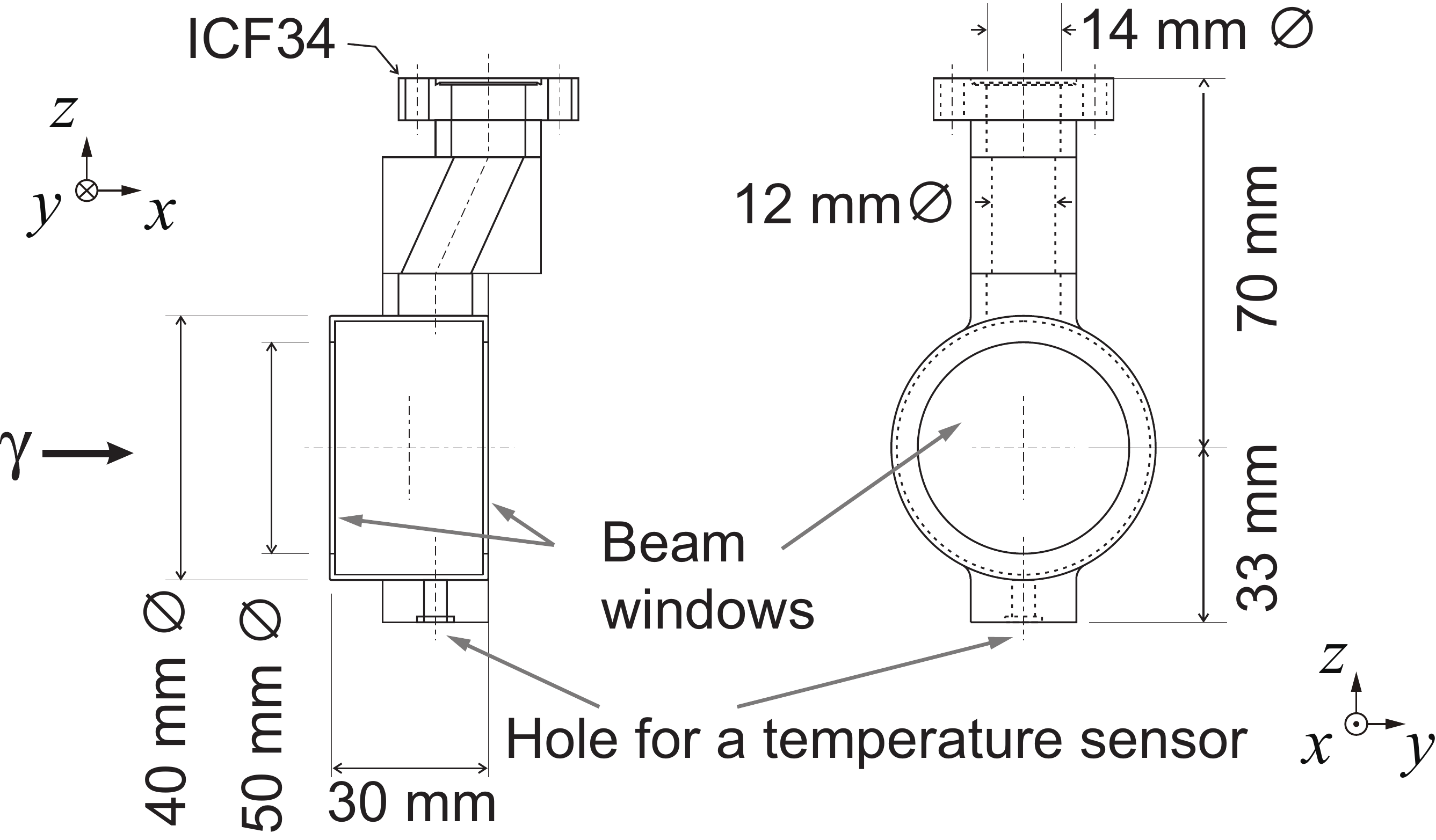}
      \caption{
			Drawing of the target cell.
			Small part at the bottom was for the insertion of the temperature sensor inside the cell.
			}
      \label{fig:tcell}
    \end{center}
  \end{figure}

  The target cell was a cylinder with its axis aligned to the beamline.
  It was made of a 1-mm-thick aluminum shell with a pipe for the connection to the gas pipe with an ICF-34 flange.
  A heat exchanger was placed along the gas pipe, and the liquefied gas was supplied to the cell.
  The thickness of the target was 30~mm, and its diameter of the target cell was 50~mm.
  It had two windows for the beam inlet and outlet (see Fig.~\ref{fig:tcell}).
  The windows were covered with 75-$\mu$m-thick polyimide (Ube Upilex-S) film sheets.
  The film sheets were glued to the fringes of the aluminum shell with epoxy resin (Emerson and Cuming Stycast 1266).
  The whole cell was covered with a cylindrical super-insulator made of an aluminum-evaporated Mylar film.

  The temperature of the liquid was measured with two temperature sensors.
  One was inserted from the bottom of the cell, and the other was placed in the gas pipe.
  Both of them were immersed in the liquid when the system reached stable state.
  The period needed to reach the stable state after starting the filling of the liquid for deuterium (hydrogen) were 4 (2) hours.

  The pressure of the gas was measured along the gas supply pipe outside of the cryostat.
  At the stable state, the measured pressures were stable and no flow of the gas was expected.
  Thus, the pressure of the gas and the liquid in the system was considered the same.
  The typical absolute pressure was 50 kPa for both deuterium and hydrogen.
  The typical temperatures of liquid deuterium (hydrogen) in the cell were 20.2 (15.3) K.
  They were sufficiently lower than the boiling point at the pressure.

  No boiling was observed during a test with a small view hole in the super-insulator. 
  The temperature of the liquid in the gas pipe was lower than the temperature at the bottom of the cell. 
  The liquid in the pipe contacted the heat exchanger, and thus the temperature was kept low.
  However, the liquid in the cell was heated by thermal radiation through the windows of the target cell.
  The typical temperature difference of the liquid deuterium (hydrogen) was 0.25 (0.1) K.
  It affected the temperature inhomogeneity and thus the density inhomogeneity of the liquid in the cell.

  The density of the liquid was evaluated from the temperature measured in the cell and the gas pressure.
  The typical density of liquid deuterium (hydrogen) was 0.172 (0.0762) g/cm$^3$. 
  The uncertainty of the density inhomogeneity of liquid deuterium (hydrogen) was considered to be 0.3 (0.1)~\%.
  The stability of the temperature of liquid deuterium (hydrogen) was 0.05 K / 6 days  (0.15 K / 3 days).
  The stability of the pressure of deuterium (hydrogen) was 1 kPa / 6 days (1.3 kPa / 3 days).
  They affected the density by 0.1 (0.2)~\%.
  The details of the target system can be found in Ref.~\cite{Kanda:2004ex}.

%%___________________________________________________________________________________________________________________

\section{Tagged Photon Beams}

  The LNS-Tohoku accelerator complex comprised an electron linear accelerator (LINAC) and a STretcher-Booster (STB) ring.
  150 MeV electrons were injected from the LINAC and accelerated up to a maximum energy of 1.2 GeV in the STB ring~\cite{Hinode:2001aa, Hama:2001aa}.
  In order to generate the photon beams, a movable carbon wire radiator of $11~\mu$m {\o} was inserted into the path of the electron beam.
  The radiator could move horizontally and perpendicularly to the electron beam axis.

\begin{figure*}[htbp]
    \begin{center}
      \includegraphics[bb=0 0 647 366,width=13.5cm,clip]{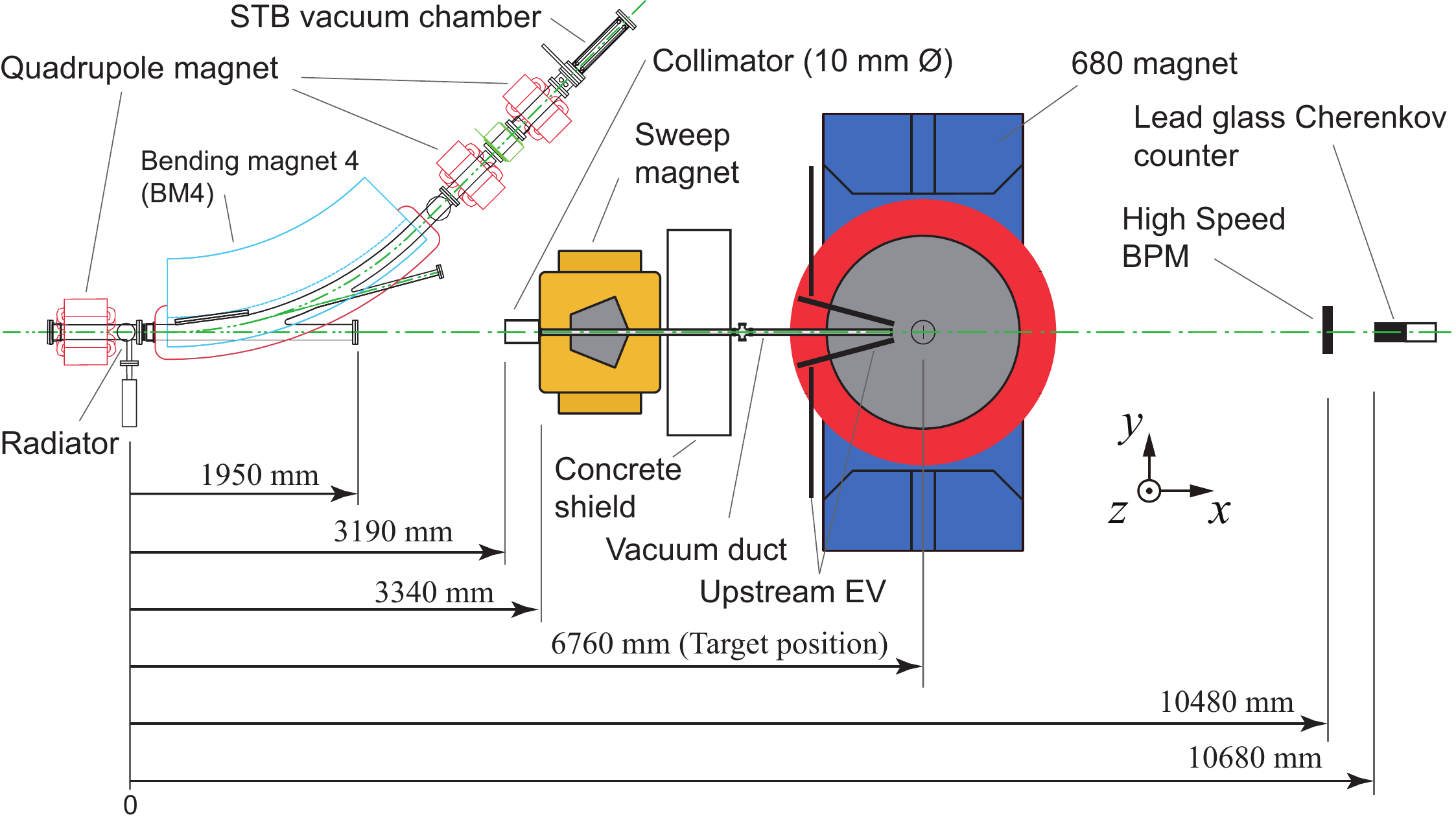}
      \caption{
			Top view of the beamline setup.
			A lead collimator, sweep magnet, and vacuum chamber were located upstream from the target.
			A high-speed beam profile monitor (HSBPM) and a lead-glass Cherenkov counter were placed downstream from the target.}
      \label{fig:beamline}
    \end{center}
  \end{figure*}

  An electron predominately loses energy by emitting a bremsstrahlung photon.
  The emitted photon is directed to NKS2, and simultaneously the recoiled electron is tagged by the STB-Tagger system.
  The schematic view of the photon beamline is shown in Fig.~\ref{fig:beamline}.
  The detailed description of the radiator control and the photon tagging system can be found in Ref.~\cite{Yamazaki:2005ex}
	
%_____________________________________

\subsection{STB-Photon Tagger}
  \label{ss:STBTagger}

  The STB tagger system was located in the gap of Bending Magnet 4 (BM4) of the STB ring.
  The components included the aforementioned carbon radiator, BM4, and two arrays of individual tagging counters named as TagF and TagB.
  TagF and TagB consist of 48 and 12 plastic scintillation counters, respectively.
  Scintillation light from each counter was transported to the PMT by an optical fiber bundle.

  TagF had the role to identify the energy of the recoil electron.
  The use of TagB in coincidence with TagF was primarily to select recoil electrons 
  after photon emission at the radiator from other sources of background such as M{\o}ller scattered electrons
  and secondary particles from the hit of electrons on the beam duct. 
  The timing information of TagF and TagB was recorded by a TDC module (CAEN V775).
  The charge information of TagB signal was recorded by a charge-to-digital converter (QDC) module (CAEN V792) to enable time walk correction to obtain a better timing selection.
  The tagger system was capable of tagged photon energies over a range of 0.8--1.1 GeV with an accuracy of $\pm 10$ MeV on the produced photon beam~\cite{Yamazaki:2005ex}.

  Calibration of the tagging detectors resulted in a correlation between the tagger segment number and the energy of a photon incident on the target;
  therefore, it provided information on the photon energy and the number of photons.
  The photon energy is deduced from energy conservation as follows:
  \begin{equation}
    E_{\gamma} = E_{e}-E_{e'}-E_{\mathrm{recoil}}, 
  \end{equation}
  where $E_{\gamma}$ is the energy of the photon, $E_{e}$ is the energy of the electron in the STB ring, 
  $E_{e'}$ is the energy of the electron as measured in the tagging system, 
  and $E_{\mathrm{recoil}}$ is the recoil energy of the radiator nucleus of bremsstrahlung.
  Because the recoil energy is negligibly small, we assumed $E_{\mathrm{recoil}} = 0$.
  In addition to that method, the kinematically complete measurement of $\gamma+d \rightarrow p+p+\pi^-$ was successfully established as a method of calibrating the photon tagging system~\cite{Yun-Cheng:2010ab}.

  The lead-glass Cherenkov (LG) counter was used to measure the number of photons that passed through the spectrometer, which is equivalent to the approximate number that reached the target.
  The lead glass was SF5, which was provided by OHARA Optical Glass Mfg. Ltd. (now, OHARA INC.).
  The tagging efficiency for each TagF was evaluated as the number of photons detected by the LG counter divided by the number of photons detected by TagF. 
  The photon beam intensity was kept at a few hertz predominantly owing to the counting rate capability of the LG counter and to reduce the probability of coincidences between the LG counter and the tagger.
  In principle, this should have no effect on the measured efficiency.
  The tagging efficiency was 75--80~\% over the TagF counters.
  The LG counter was prepared for the FOREST detector, and the performance study is described in Ref.~\cite{Ishikawa:2016kin}.

%___________________________________________________

\subsection{Sweep Magnet}

  A large number of $e^+ e^-$ pairs created upstream from the photon beam was substantially reduced by the sweep magnet ($B = 1.1$~T at I = 300~A).
  In front of the sweep magnet, there was a collimator comprised of five lead blocks (250 mm thickness in total) to reduce the beam halo.
  The collimator aperture was 10~mm~{\o}.
  The sweep magnet, being located before the main spectrometer, 
  efficiently suppressed the background contribution to the data and improved the DAQ rate.

  However, electrons and positrons from upstream were not completely removed by the sweep magnet,
  and thus two sets of EV counters were placed upstream of NKS2 at the same height of the photon beams 
  in order to reject them in the trigger (see Section~\ref{sec:ev}).
 
%___________________________________________________

\subsection{Beam Profile Monitor}

  A high-speed beam profile monitor (HSBPM) was composed of two layers of scintillating fiber bundles.
  Each layer had 16 scintillating fibers (Saint-Gobain, BCF-10SC with black extra mural absorber coating) of 3-mm square cross-section
  and read out by a 16-ch multianode PMT (HAMAMATSU Photonics, H6568-10).
  One bundle was horizontally aligned, and the other was vertically aligned.
  They crossed over a $48 \times 48$ mm square region to provide two-dimensional hit information by the coincidence of the vertical 
  and horizontal channels for charged particles.
  Charged particles were provided from the photon beams by an aluminum converter plate having a thickness of 0.1 mm ($\sim 1\times10^{-3} X_0$).
  It also consisted of a pair of trigger counters and a veto counter to ensure that electrons and positrons converted from photons generated the trigger.
  The beam profile was checked by HSBPM when the beam course was tuned in the experimental period.
  The detailed information can be found in Refs.~\cite{Nanao:2003ex, Ishikawa:2010aa}.

%___________________________________________________

\subsection{Electron Beam Structure}

  A typical beam cycle consisted of times of waiting, beam injection, ramping up, storage (flat top), and ramping down.
  During the storage time, the radiator was inserted into an orbit of electrons, and the photon beams were impinged on the target.
  The duty factor (DF) was defined as the ratio of the flat-top time to the period of a single cycle.

  The flat-top and waiting times could be changed upon user request and were limited by power consumption.
  The time for ramping both up and down was established at 1.4 s.
  Typically, the DF was approximately 75~\% at the period of 53 s and the flat top of 40 s.

%%___________________________________________________________________________________________________________________
   
\section{Trigger and DAQ}

  We applied a minimum bias trigger for two-charged-particle event in $\gamma+d$ or $\gamma+p$ reactions.
  In the previous NKS experiment, we adopted the upstream and downstream EV counters in the trigger, 
  which rejected all charged particles on the beam plane to reduce the trigger rate.
  However, that introduced a bias.
  Its effect was especially large in the downstream direction.
  We avoided that trigger bias by excluding the downstream EV counter in the trigger, but the upstream EV counter was included as described in Section~\ref{sec:ev}.

%___________________________________________________

\subsection{Trigger}

  The trigger was formed from the combination of the photon tagger, IH multiplicity, OH multiplicity, and upstream EV.
  The upstream EV counter had a role to reject $e^{+}e^{-}$ events produced in the upstream region. 
 
  The trigger logic is shown as:
  \begin{equation}
    \label{eq:trigger}
    \sum_{i=1}^{12} [TagH_{i} \otimes (nIH \geq 2)]
    \otimes
    (nOH \geq 2)
    \otimes
    \overline{(nEV \geq 1)}
  \end{equation}

  The photon tagging system had 12 units.
  $TagH_{i}$ was made from the coincidence of one TagF hit and TagB hit in the $i$-th unit.
  $nIH$ and $nOH$ were the multiplicities of IH hits and OH hits, respectively.
  In the experimental hall, each counter hit of IH and OH was defined as the coincidence signal of up and down (or upstream and downstream) PMTs.
  $nEV$ was taken as a logic-OR signal of the upstream EV counter, which was located the upstream side of the beam plane to reject $e^{+} e^{-}$ background.

  The multiplicity logic was implemented with a special VME module (named TUL-8040) using field-programmable gate arrays (FPGA, Altera APEX 20K300EQC240-1X).
  The maximum internal clock of TUL-8040 was 300 MHz.
  We originally developed TUL-8040 for the ($e, e' K^{+}$) hypernuclear spectroscopy experiment (JLab E01-011 and E05-115)~\cite{Nakamura:2012hw, Tang:2014atx}.
  We also used TUL-8040 to make the mean time of a OH hit, computed from the timing information of the top and bottom (upstream and downstream) sides of the PMT  for  OHV (OHH).

  Each detector trigger was generated in the second experimental hall.
  These trigger signals were sent to the counting room, and the master trigger shown by formula \eqref{eq:trigger} was generated from them.
  The typical trigger rate was 1 kHz at a tagger rate of 2.5 MHz.

%___________________________________________________

\subsection{DAQ System}

  Data were taken by using three Linux PCs.
  One PC was assigned to record signals from hodoscopes and taggers, and the other two PCs were assigned to record drift chamber signals.
  UNIDAQ~\cite{Nomachi:1994tk} was installed as a DAQ system.
  The event building was performed in offline analysis.
  A four-bit increasing number was generated by TUL-8040 and distributed to I/O register modules (REPIC RPV-130).
  We adopted CAEN VME modules: V775 TDC modules and V792 QDC modules for counter signals.
  As described in Section~\ref{sec:DC_readout}, AMT-VME modules were used for the timing measurement of drift chamber hits. 

  The numbers of requested triggers, accepted triggers, and hits on some counters were recorded by CAMAC scalers.
  The scaler values were read at the end of the beam spill and then reset to zero.
  The spill period was given by counting a signal of clock generator between the radiator start and stop signals.
  Those signals are distributed from the radiator controller (see Ref.~\cite{Yamazaki:2005ex} in detail).
  
  We used a PCI-VME adaptor, SBS Technologies (Bit3) Model 620-3, to control VME modules.
  The TDC and ADC information was read event by event.
  The CAMAC scaler data were collected via a CAMAC controller (Toyo Corporation, CC-7700).

%%___________________________________________________________________________________________________________________
 
\section{Performance of the NKS2 System}

\subsection{Time Resolution}

  Time resolution of the hodoscopes was evaluated by using electrons and positrons, whose velocities can be assumed to be $c$ ($\beta = 1$).
  The hit timing of each counter was corrected by taking the time walk into account.
  The time walk correction for each counter was performed by using the pulse charge information recorded by the QDC module.
  Relative timing among IH counters was adjusted to be zero, assuming that the deviation of the track lengths from the reaction point to IH counters was negligibly small.
  Relative timing between IH and OH was calibrated for the TOF of electrons and positrons using the flight lengths obtained by tracking using the drift chambers.

  For the evaluation of an intrinsic time resolution of each counter, 
  we estimated the spreads of time differences between IHL2 to IHR2, IHL2 to OHVL2, IHR2 to OHVR2, IHL2 to OHVR2, and IHR2 to OHVL2.
  They were denoted as $\sigma_{\mathrm{IHL2-IHR2}}$, $\sigma_{\mathrm{IHL2-OHVL2}}$, $\sigma_{\mathrm{IHR2-OHVR2}}$, 
  $\sigma_{\mathrm{IHL2-OHVR2}}$, and $\sigma_{\mathrm{IHR2-OHVL2}}$, respectively.
  They could be expressed with intrinsic time resolutions of relevant counters,
  $\sigma_{\mathrm{IHL2}}$, $\sigma_{\mathrm{IHR2}}$, $\sigma_{\mathrm{OHVL2}}$, $\sigma_{\mathrm{OHVR2}}$,
  and an imperfectness of time adjustment, $\sigma_{\mathrm{adj}}$.

%  \begin{strip}
  \[
    \begin{array}{ccccccc}
		\sigma_{\mathrm{IHL2-IHR2}}^2  & = & \sigma_{\mathrm{IHL2}}^2 & +\sigma_{\mathrm{IHR2}}^2 &                            &                            &                          \\
		\sigma_{\mathrm{IHL2-OHVL2}}^2 & = & \sigma_{\mathrm{IHL2}}^2 &                           & +\sigma_{\mathrm{OHVL2}}^2 &                            &                          \\
		\sigma_{\mathrm{IHR2-OHVR2}}^2 & = &                          &  \sigma_{\mathrm{IHR2}}^2 &                            & +\sigma_{\mathrm{OHVR2}}^2 &                          \\
		\sigma_{\mathrm{IHL2-OHVR2}}^2 & = & \sigma_{\mathrm{IHL2}}^2 &                           &                            & +\sigma_{\mathrm{OHVR2}}^2 & +\sigma_{\mathrm{adj}}^2 \\
		\sigma_{\mathrm{IHR2-OHVL2}}^2 & = &                          &  \sigma_{\mathrm{IHR2}}^2 & +\sigma_{\mathrm{OHVL2}}^2 &                            & +\sigma_{\mathrm{adj}}^2 \\
   \end{array}
  \]
%  \end{strip}

\begin{figure}[htbp]
    \begin{center}
      \includegraphics[bb=0 0 346 495,width=7.5cm]{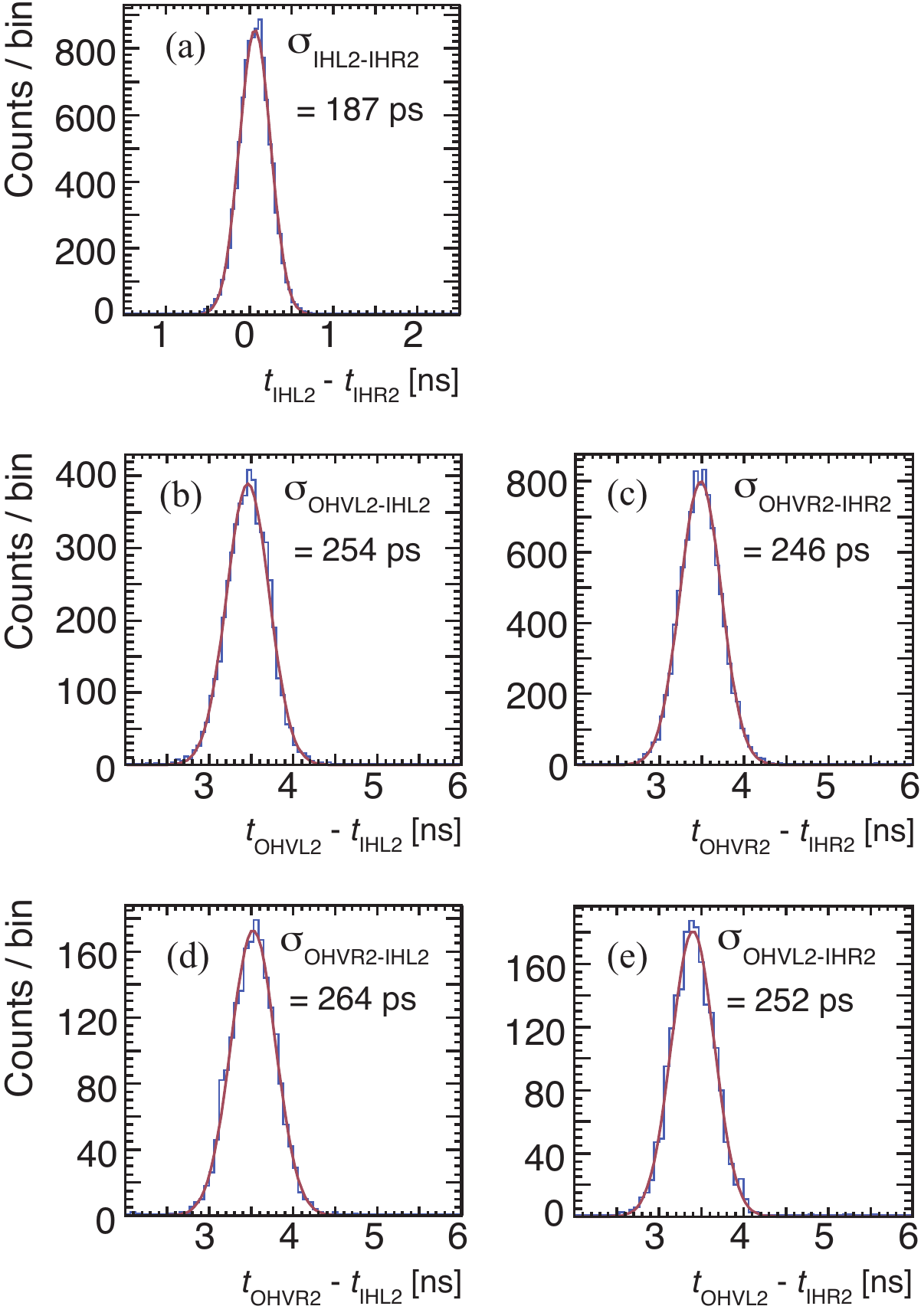}
      \caption{
				Time difference distributions for (a) IHL2 to IHR2, (b) IHL2 to OHVL2, (c) IHR2 to OHVR2, (d) IHL2 to OHVR2, and (e) IHR2 to OHVL2.
				They were fitted by the Gaussian and the obtained $\sigma$s are shown in the panels.
              }
      \label{fig:timespread_IH_OH}
    \end{center}
  \end{figure}

  Figure~\ref{fig:timespread_IH_OH} shows the time difference between IH and OH counters.
  The relative time spreads were obtained by the Gaussian fit of the time difference distributions.
  Solving the set of equations above and substituting the obtained relative time spreads, the intrinsic time resolutions were obtained:

  \begin{eqnarray*}
    \sigma_{\mathrm{IHL2}}  & = & 143~\mathrm{ps} \\
    \sigma_{\mathrm{IHR2}}  & = & 121~\mathrm{ps} \\
    \sigma_{\mathrm{OHVL2}} & = & 204~\mathrm{ps} \\
    \sigma_{\mathrm{OHVR2}} & = & 205~\mathrm{ps} \\
    \sigma_{\mathrm{adj}}   & = &  87~\mathrm{ps}
  \end{eqnarray*}
 
  Using those intrinsic time resolutions and time spreads between the counters, we obtained average intrinsic time resolutions.
  Those were 140 ps for IH counters and 200 ps for OH counters as $\sigma$ of the Gaussian.

\begin{figure}[htbp]
    \begin{center}
      \includegraphics[bb=0 0 379 283,width=6.5cm, clip]{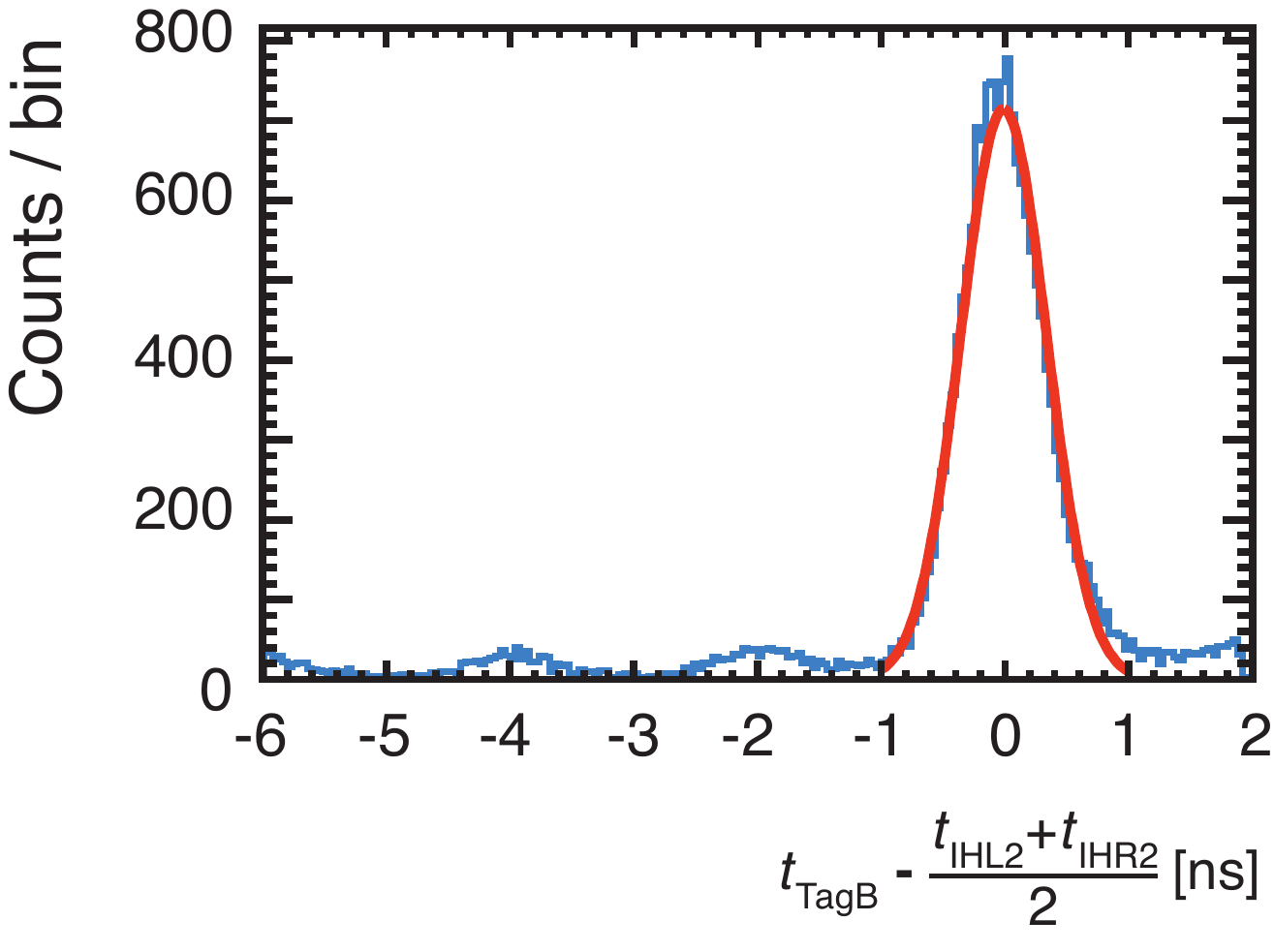}
      \caption{
               Time difference between IH and tagger for $e^+ e^-$ events. 
               The periodic time structure of 2 ns is caused by the microbunch structure of the electron beam.
               The width of the true coincident peak is 350 ps ($\sigma$).
              }
      \label{fig:timespread_IH_TagB}
    \end{center}
  \end{figure}

  We utilized the timing information on the tagging counter of the STB tagger for the event selection.
  TagBs described in section \ref{ss:STBTagger} were used to define the electron hit timing on the STB tagger.
  The time difference between the averaged IH and averaged TagB ($t_{\mathrm{TagB}} - (t_{\mathrm{IHL2}} + t_{\mathrm{IHR2}})/2$) is shown in Fig.~\ref{fig:timespread_IH_TagB}.
  A clean peak around 0 ns corresponds to true coincidence events. 
  The smaller peaks at $-2, -4$ ns are caused by accidental coincidences, which reflected the microbunch structure of the electron beam (radio frequency of the accelerator was 500.14 MHz ($\simeq$ (2 ns)$^{-1}$) ).

  The time spread of the true coincident peak was obtained as 350 ps ($\sigma$).
  Using the time resolution of IHs, the intrinsic time resolution of TagB was estimated to be 340 ps ($\sigma$). 

%__________________________

\subsection{Position Resolution}
  \label{sec:position_resolution}

  We estimated the position resolutions of VDC and CDC with a help of a GEANT4 simulation.
  In the process of drift length calibration, we obtained the residual distribution between the ``drift length computed from the drift time" and the ``distance of the closest approach (DCA) between the sense wire and the track."
  The residual distribution depends on the position resolutions of the drift chambers.
  We assumed some combinations of the position resolutions for VDC and CDC in the simulation and calculated the residual distribution.
  The simulation with the position resolutions of VDC $\sigma = 450$ $\mu$m and CDC $\sigma = 350$ $\mu$m reproduced the experimental data well.

%___________________________

\subsection{Momentum Resolution}

  We reconstructed the trajectories of charged particles from VDC and CDC information by a fitting method.
  To compute the trajectory in the magnetic field, we adopted the Runge--Kutta method introduced in Ref.~\cite{Myrheim:1979ng}.
  The detailed formula of the full 3D Runge--Kutta method is given in \ref{appendix:runge_kutta}; the formula was not explicitly shown in the reference.

  The equation of motion is given as a differential equation of time ($t$) in the Cartesian coordinate space ($x, y, z$).
  We rewrite the equation of motion as shown in Ref.~\cite{Wind:1974zz}.
  That process gives us the simultaneous differential equations about $y'(=dy/dx)$ and $z'(=dz/dx)$. 

\begin{figure}[htbp]
    \begin{center}
      \includegraphics[bb=0 0 467 315,width=7.5cm, clip]{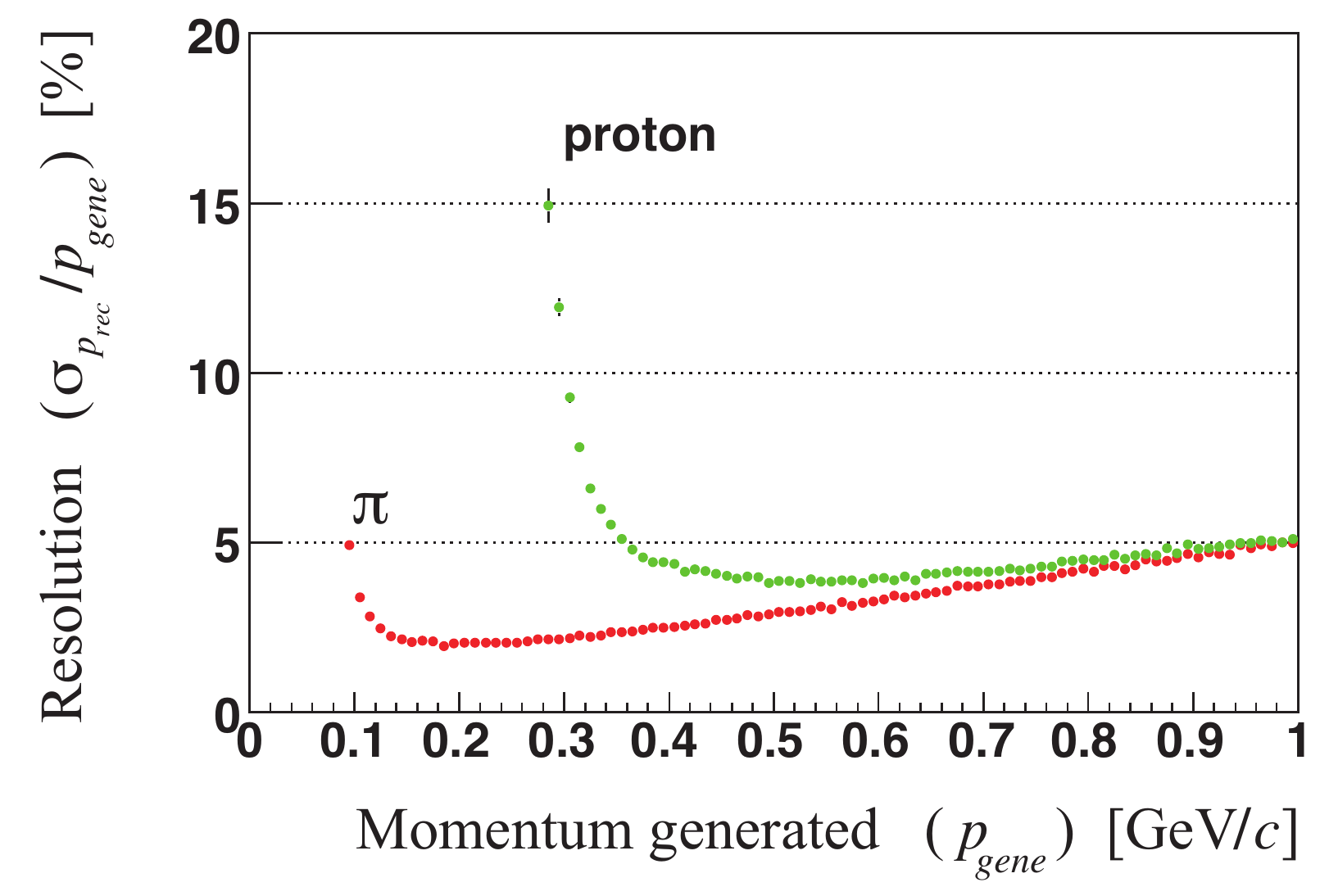}
      \caption{
			Relative momentum resolution estimated by a Monte Carlo simulation using GEANT4 as a function of the momentum of generated particles.
			$\sigma_{p_{rec}}$ is a standard deviation of the Gaussian obtained by fitting to the momentum reconstructed.
			The momentum of generated particles is denoted by $p_{gene}$. 
              }
      \label{fig:momentum_resolution}
    \end{center}
  \end{figure}

  The resolution was estimated by the GEANT4 simulation. 
  Figure~\ref{fig:momentum_resolution} shows the relative momentum resolution ($\sigma_{p_{rec}}/p_{gene}$) for generated pions and protons as a function of momentum.
  The relative momentum resolution was less than 5~\% for 0.1 to 1.0 GeV/$c$ of the pion momentum and 0.35 to 1.0 GeV/$c$ of the proton momentum when we used both VDC and CDC hits.
  Those values are sufficient for $K^{0}_{S}$ and $\Lambda$ identification by invariant mass and the cross-section measurement.

  We also studied the effect of the position resolution on the invariant mass resolution to check the consistency, 
  because we did not measure the momentum resolution in the experiment directly.
  See Section~\ref{sec:invariant_mass} and \ref{sec:3_track_analysis} for the comparison with the data.

  The current tracking algorithm has no correction for the energy loss effect,
  and thus a Monte Carlo simulation shows that the momentum of reconstructed tracks systematically shifts to lower values.
  This effect was taken into account in the acceptance correction when we computed the cross-section as a function of momentum.
  We plan to adopt the Kalman filter method when we need higher precision of momenta.

%_______________________________________

\subsection{Particle Identification}

  Particle identification was carried out by its mass and the charge.
  The mass $m$ was calculated with the momentum $p$ and the velocity $\beta$:
  \begin{equation}
    m^2= p^2 \left( \frac{1}{\beta^2} -1  \right).
    \label{eq:7_mass}
  \end{equation}
  The sign of charge was given by the bending direction in the magnetic field.

  \begin{figure}[thb]
    \begin{center}
      \includegraphics[bb=0 0 565 533,width=7.5cm]{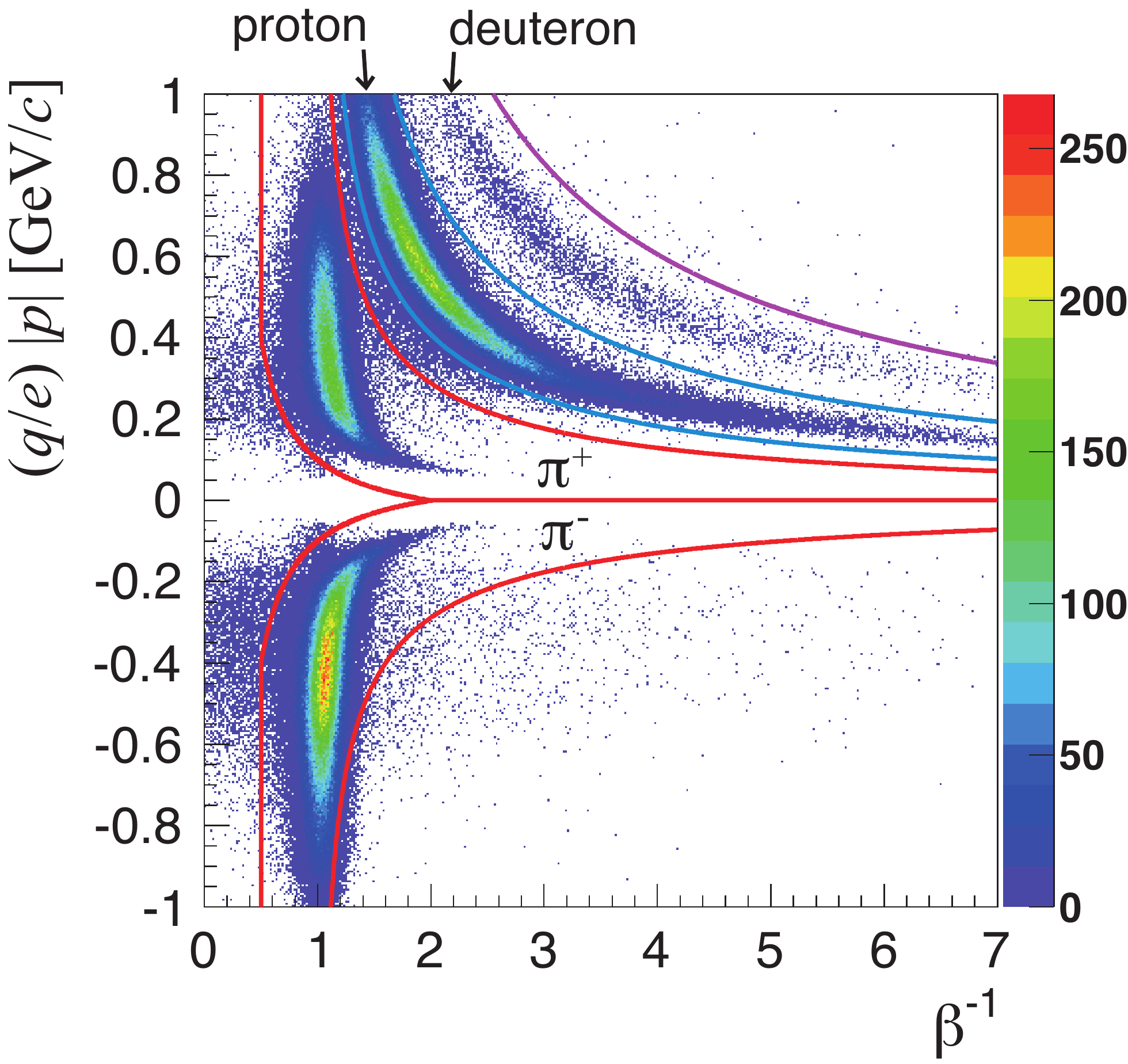}
      \caption{
			Scatter plot of the momenta and the inverse velocities for charged particles. 
			The sign of the vertical axis represents the charge sign of the particle. 
			The opening angles ($\eta$) of the two tracks were selected by the condition of $-0.9 < \cos \eta < 0.8$ to remove $e^+ e^-$ events.              }
      \label{fig:pid}
    \end{center}
  \end{figure}

  Figure~\ref{fig:pid} shows the scatter plot of the momentum and the inverse of the velocity of particles.
  The opening angle ($\eta$) selection with $-0.9 < \cos \eta < 0.8$ was applied in order to reject $e^{+}e^{-}$ photon conversion events.
  The sign of the vertical axis expresses the charge sign of the particle.
  The conditions of the particle identification for the pion were defined as:
  \begin{equation*}
    \left.
    \begin{aligned}
        0.5 \leq \beta^{-1}, \\
        |p| < \frac{0.1444}{\beta^{-1} -0.2} -0.08
    \end{aligned}
    \right\}
       ~ \mathrm{for}~ \beta^{-1} \leq 2.0,
  \end{equation*}
  \begin{equation}
     -0.5 < m^2 < 0.25 ~~ \mathrm{for}~ 2.0 \leq \beta^{-1}.
    \label{eq:mass_pion}
  \end{equation}
  The $m^{2}$ condition for proton was:
  \begin{equation}
    0.5 < m^2 < 1.8.
    \label{eq:mass_proton}
  \end{equation}
  It for deuteron was:
  \begin{equation}
    1.8 < m^2 < 5.5 .
   \label{eq:mass_deuteron}
  \end{equation}
  The units of mass and momentum are GeV/$c^2$ and GeV/$c$, respectively.

  The selected regions for the pion, proton, and deuteron in the $p$ - $\beta^{-1}$ scatter plot are shown in Fig.~\ref{fig:pid}.
  The areas separated by lines correspond to the select conditions derived by Eqs.~\eqref{eq:mass_pion},~\eqref{eq:mass_proton},  and~\eqref{eq:mass_deuteron}.

%___________________________________________________

\subsection{Decay Vertex Finding and Its Resolution}

\begin{figure}[htbp]
    \begin{center}
      \includegraphics[bb=0 0 560 653,width=7.5cm]{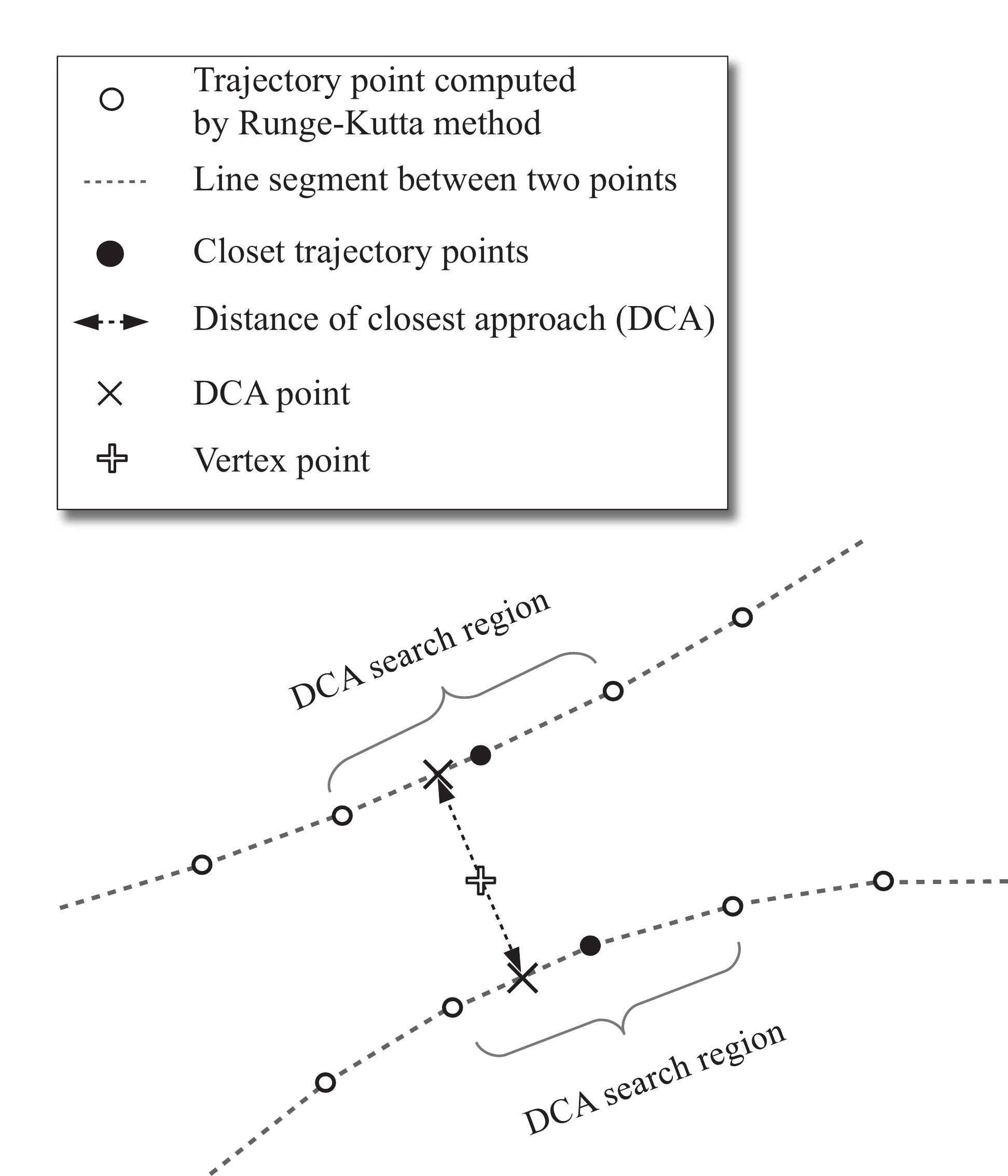}
      \caption{
				Illustration of decay vertex finding.
				See the text for the algorithm.
			}
      \label{fig:vertex_search}
    \end{center}
  \end{figure}

  By combining positively and negatively charged tracks, we found the decay vertex position and reconstructed the energy and momentum of the parent particle.
  The vertex finding was performed by the following steps.
  \begin{enumerate}
	\item choose a combination of positively and negatively charged tracks from the list of reconstructed tracks.
	\item find the closest trajectory points from two tracks (solid circle in Fig.~\ref{fig:vertex_search}).
	\item select the closest line segments (dashed line in Fig.~\ref{fig:vertex_search}) from each trajectory in the DCA search region (indicated by curly brackets in Fig.~\ref{fig:vertex_search}).
	\item obtain DCA points (X in Fig.~\ref{fig:vertex_search}) on the line segments and calculate the center of them as the vertex point.  
  \end{enumerate}

\begin{figure}[htbp]
    \begin{center}
      \includegraphics[bb=0 0 516 415,width=7.5cm]{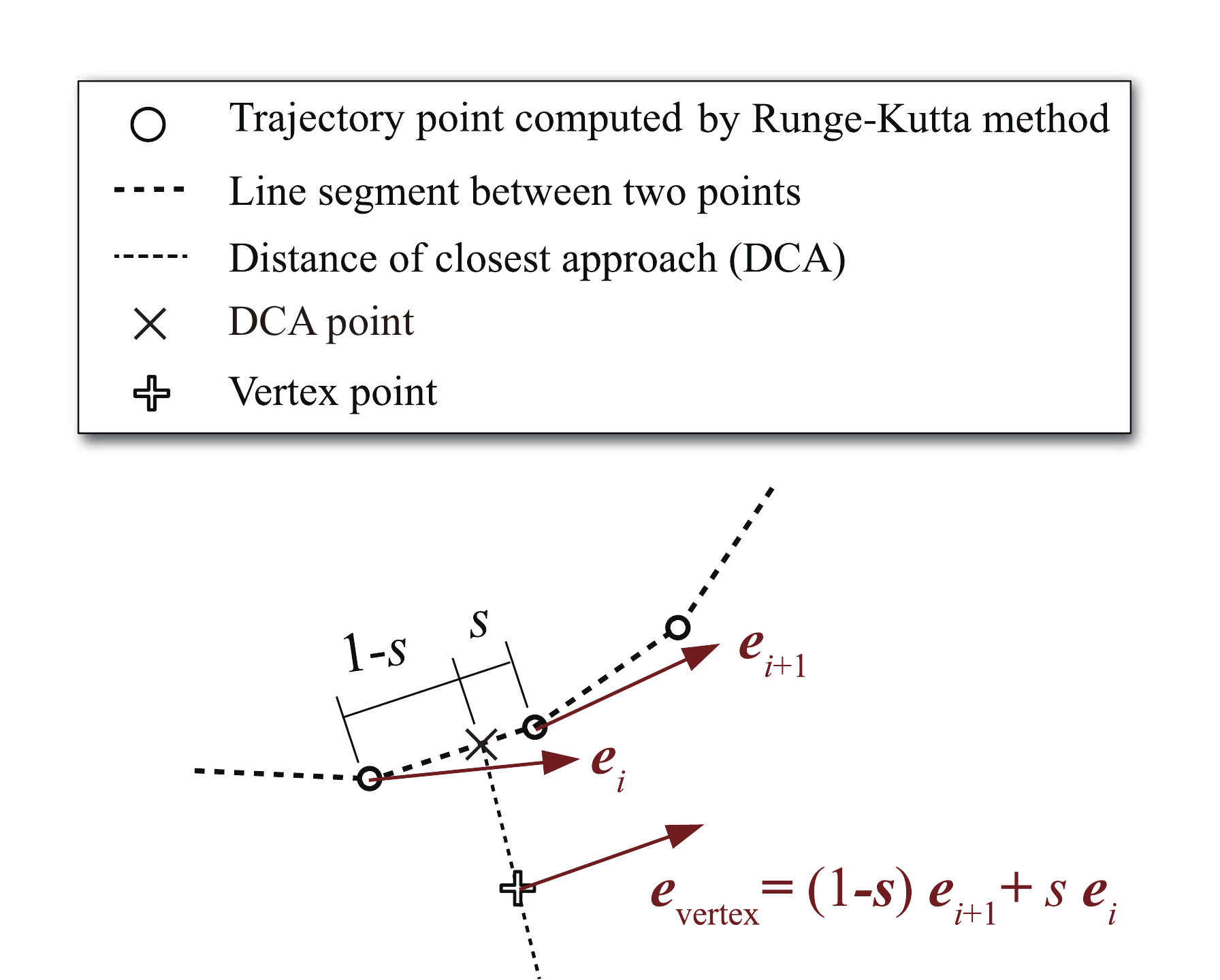}
      \caption{
				Illustration of how to estimate the momentum at the decay vertex.
				One side of the trajectory is demonstrated.
				$\boldsymbol{e}_{i}$ and $\boldsymbol{e}_{i+1}$ are the unit vectors of the momentum direction at trajectory points given by the Runge--Kutta method.
				The DCA points are on the line segment between the $i$-th and $(i+1)$-th trajectory points.
				The fraction $s$  $(0{\le}s{\le}1)$ and $1 - s$ are used as weights to evaluate the direction vector at the vertex.
				The momentum at the decay vertex is given by multiplying the absolute momentum to $\boldsymbol{e}_{\mathrm{vertex}}$.  
			}
      \label{fig:momentum_at_vertex}
    \end{center}
  \end{figure}

  The Runge--Kutta calculation routine recorded the momentum vector at each point of the steps.
  The momentum vector of each track at the vertex point was given as:
  \begin{enumerate}
	\item obtain two momentum direction vectors at both edges of the line segment where the DCA point exists.
	\item take a weighted average of the direction vectors, 
			where the weight is a fraction of ``the distance between the DCA points to the edge" to the length of ``the line segment" (see Fig.~\ref{fig:momentum_at_vertex}).
  \end{enumerate}
  The particle mass in four-momentum of a daughter particle was assigned as the Particle Data Group (PDG)~\cite{pdg} value after particle identification with momentum and $\beta^{-1}$.
  The four-momentum of the parent particle was calculated from those of daughter particles.

\begin{figure}[htbp]
    \begin{center}
      \includegraphics[bb=0 0 472 361,width=7.5cm]{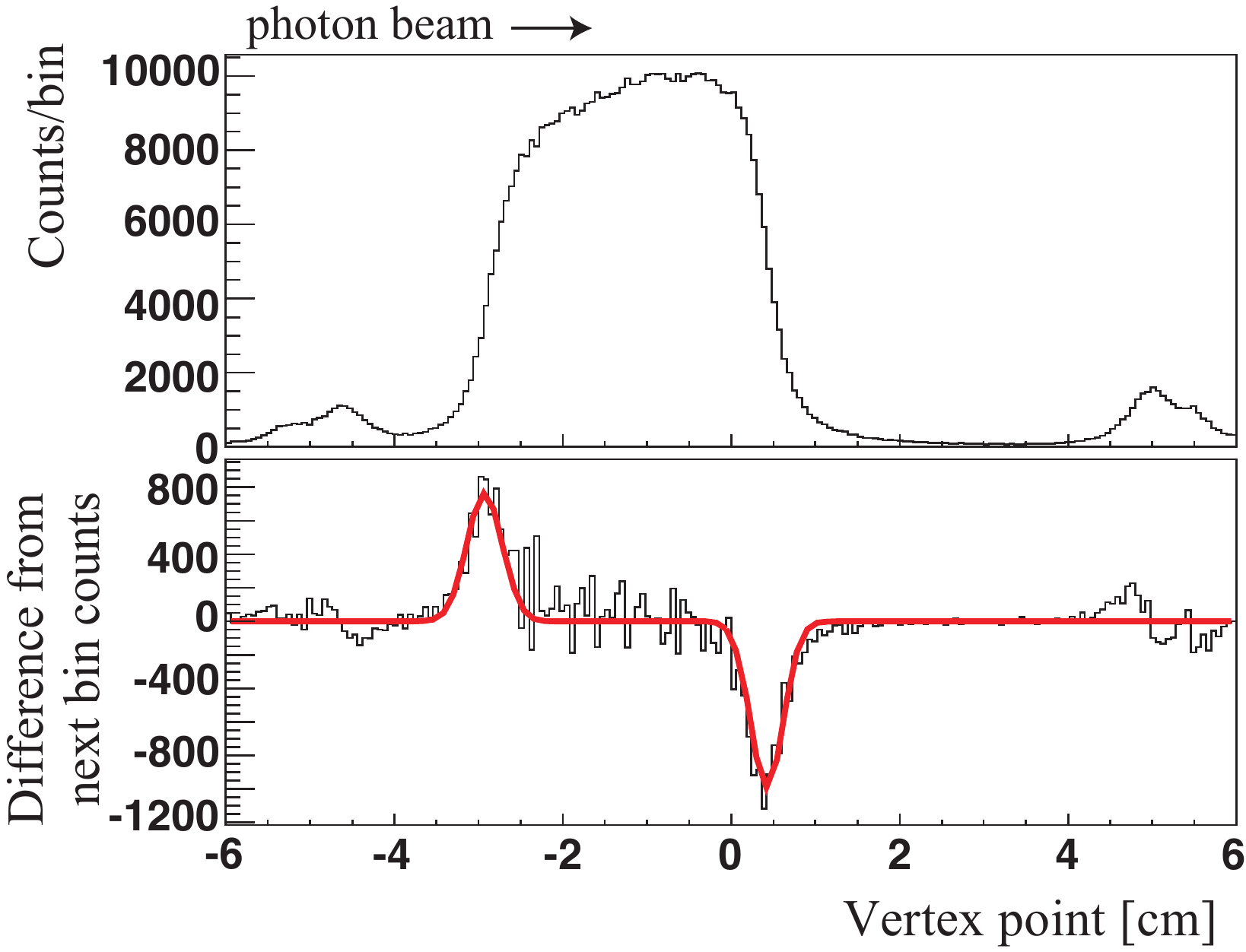}
      \caption{
			Vertex distribution of $\pi^ {+} \pi^{-}$ pairs in the beam direction is shown in the top figure.
			The bottom figure shows the difference of counts between in a bin and the next one to evaluate the vertex resolution.
			Curve in bottom figure is the Gaussian fit to evaluate the vertex resolution at the film position.
              }
      \label{fig:vertex}
    \end{center}
  \end{figure}

  The vertex resolution was evaluated by using a $\pi^{+} \pi^{-}$ pair, which was expected from the resonance decay (e.g., $\rho$, $\omega$, or $\eta$).
  Figure~\ref{fig:vertex} shows the distributions of the vertex position and the difference of counts.
  The position around $-2.9$ cm and 0.4 cm correspond to a Upilex-S film having thicknesses of 75 $\mu$m (beam windows are shown in Fig.~\ref{fig:tcell}).
  There are two small peaks around $-5$ cm and 5 cm, which are the super-insulator and Upilex-S film windows of the vacuum chamber of the target system.
  The resolution was 0.2 cm ($\sigma$) by the Gaussian fit.

  The target cell thickness is 3 cm under the normal condition, but the fitting result shows 3.3 cm.
  The difference means that the beam windows of the cell were inflated in the vacuum chamber.
%  We measured the target cell size by using a laser survey meter and confirmed the measurement result was consistent with the fitting.
  We measured the inflation of the film from the inner pressure of a test cell at room temperature with a height gauge.
  We evaluated the inflation at 20 K in consideration of the temperature dependence of the elastic modulus of the film.
  The result was consistent with the target thickness measured with the vertex distribution~\cite{Kanda:2004ex}.

  The decay vertex point and the momentum of the parent particle were used for event selections.
  The reaction vertex was reconstructed from the momenta of parent particles (e.g., $K^{0}_{S}$ and $\Lambda$) for the exclusive event.
  We selected events that a parent particle passed through the target to reduce accidental background
  for the inclusive event (single $K^{0}_{S}$ or single $\Lambda$ event).
  We reported the results of the inclusive event only because of statistics. 

\begin{figure*}[htbp]
    \begin{center}
      \includegraphics[bb=0 0 133 79,width=7.5cm]{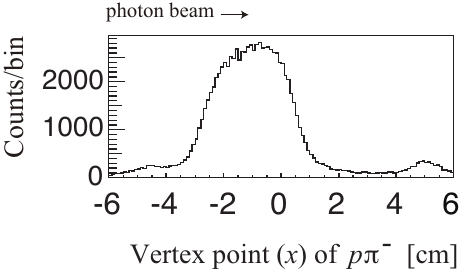}
      \caption{
 			Decay vertex distribution of $p\pi^{-}$ in the beam direction.
			}
      \label{fig:vertex_p_piminus}
    \end{center}
  \end{figure*}

  The $p\pi^{-}$ vertex distributions are shown in Fig.~\ref{fig:vertex_p_piminus}.
  The distribution along the beam axis shows longer tail in the downstream direction than the $\pi^{+} \pi^{-}$ vertex, because contributions from in-flight decay particles and resonances differ between the $\pi^{+} \pi^{-}$  and $p \pi^{-}$ events.
 
%_______________________________________

\subsection{Invariant Mass}
  \label{sec:invariant_mass}

  We reconstructed the invariant mass from positively and negatively charged particles to identify $K^0_{S}$ and $\Lambda$ with finding decay vertex points from tracking.
  To remove $e^{+}e^{-}$ background, we requested a cut of the opening angle between two tracks.
  Additionally, we applied a DCA cut between two tracks (typically less than 5 cm for $K^{0}_{S}$ and 2 cm for $\Lambda$) and requested that the reconstructed momentum of a parent particle should pass through the target volume to avoid accidental background.

\begin{figure}[htbp]
    \begin{center}
      \includegraphics[bb=0 0 115 137,width=7.5cm]{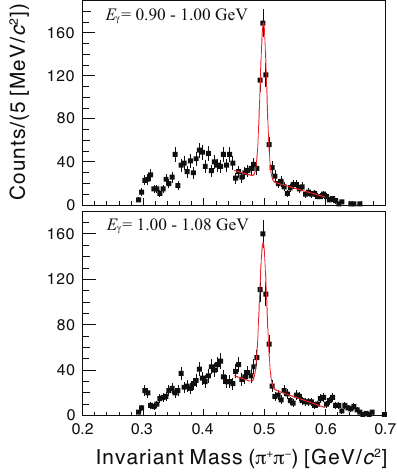}
      \caption{
                      Data point shows the invariant mass distribution from $\pi^{+} \pi^{-}$ pairs for two energy bins.
                      The curves are the fitting results of Gaussian with background of a linear function to estimate the invariant mass resolutions. 
                      The resolutions are 5.2 $\pm$ 0.3 MeV/$c^{2}$ and 5.8$\pm$ 0.4 MeV/$c^{2}$ for $E_{\gamma}$ = 0.90$-$1.00 and 1.00$-$1.08 GeV, respectively.
                   }
      \label{fig:inv_mass_pipi}
    \end{center}
  \end{figure}

  Figure~\ref{fig:inv_mass_pipi} shows the $\pi^{+}\pi^{-}$ invariant mass distributions.
  Because huge number of $\pi^{+} \pi^{-}$ events from resonance decays exist and those decay point are in the target cell, we selected events with the vertex point out of the target cell as $K^0_{S}$ candidates.

\begin{figure}[htbp]
    \begin{center}
      \includegraphics[bb=0 0 532 338,width=7.5cm]{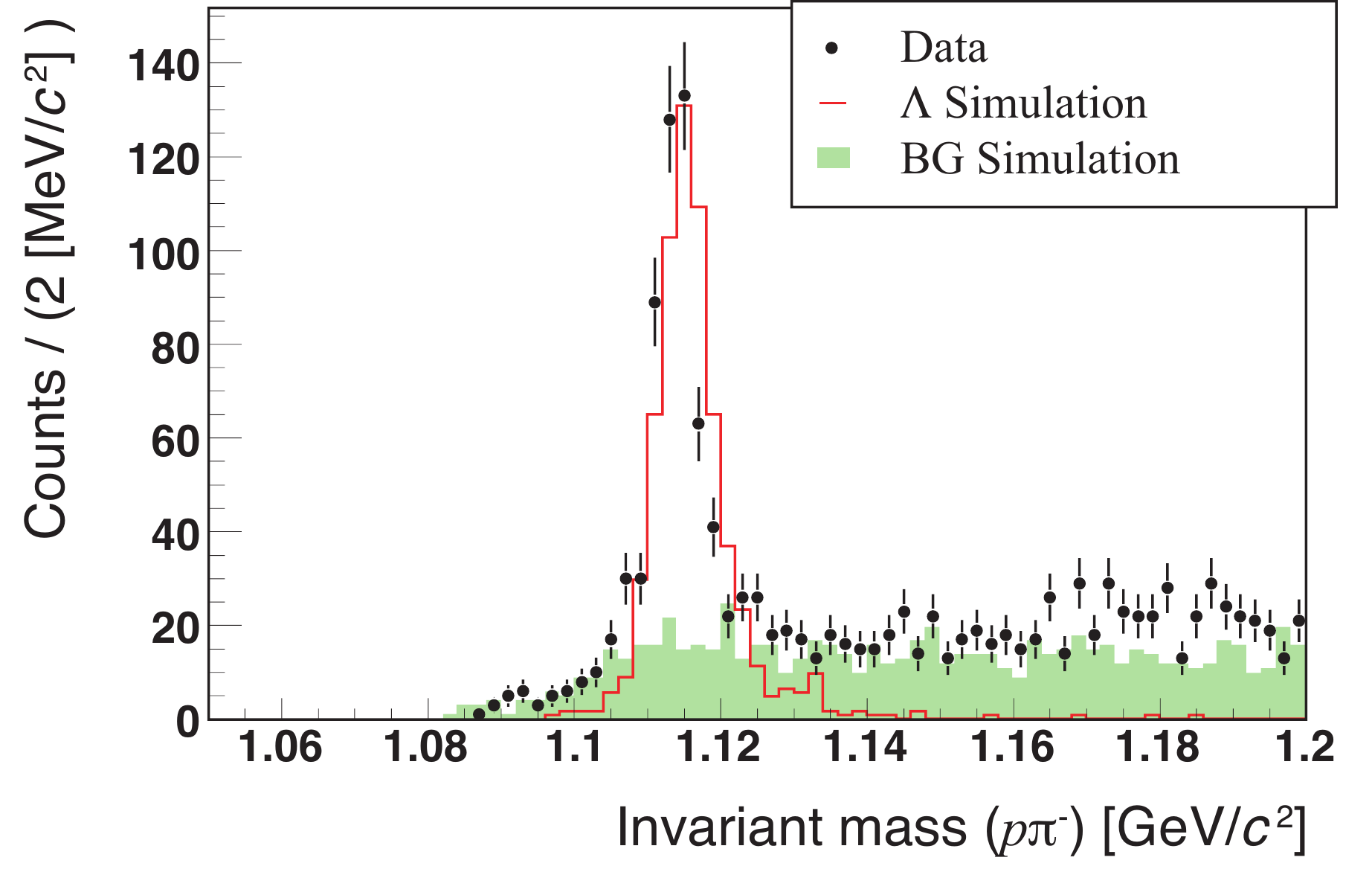}
      \caption{
                      Data point shows the invariant mass distribution from $p \pi^{-}$ pairs for $E_\gamma$ = $0.9 - 1.1$ GeV.
                      The distribution of $\Lambda$ and background were estimated with the GEANT4 simulation.
                      The mass resolution is 2.9 $\pm$ 0.2 MeV/$c^{2}$ from the Gaussian fit.
                   }
      \label{fig:inv_mass_ppi}
    \end{center}
  \end{figure}

  Figure~\ref{fig:inv_mass_ppi} shows the $p\pi^{-}$ invariant mass distributions.
  The $\Lambda$ distribution estimated by the GEANT4 simulation with the same analysis is also shown.
  The experimental data was reasonably reproduced by the simulation. 
  The invariant mass resolutions satisfied our requirement of a few MeV/$c^{2}$ for $\Lambda$ and several MeV/$c^{2}$ for $K^{0}_{S}$.

  The differential cross-section for the inclusive measurement of $\Lambda$ was calculated from the $\Lambda$ yield with corrections for the acceptance and efficiencies of the detectors and analysis.
  The results and the detailed description of the analysis can be found in Ref.~\cite{Beckford:2016aa}.

%_______________________________________

\subsection{Acceptance}

\begin{figure*}[htbp]
    \begin{center}
      \includegraphics[bb=0 0 456 154,width=12cm]{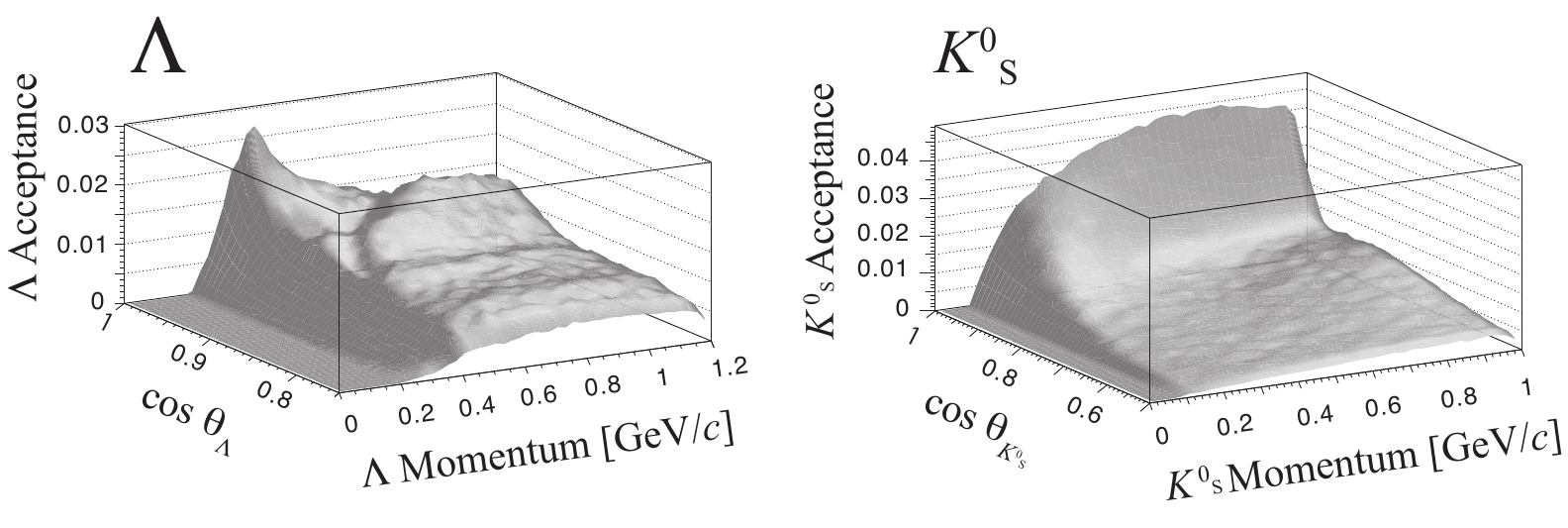}
      \caption{
				Acceptance plots of $\Lambda$ (left) and $K^0$ (right) as a function of momentum and $\cos \theta$ in the laboratory frame are shown,
				where $\theta$ is the angle with respect to the beam direction.
                   }
      \label{fig:acceptance}
    \end{center}
  \end{figure*}

  The geometrical acceptances of $K^{0}_{S}$ and $\Lambda$ were computed by the GEANT4 simulation.
  The decay modes of $\pi^{+} \pi^{-}$ for $K^{0}_{S}$ and $p \pi^{-}$ for $\Lambda$ were used in the simulation.
  Figure~\ref{fig:acceptance} shows the acceptance as a function of momentum and $\cos \theta$ in the laboratory frame, where $\theta$ is the angle with respect to the beam direction.
  Covering a forward region, the acceptance of $K^{0}_{S}$ for NKS2 is about 4 times larger than that for the previous NKS~\cite{Tsukada:2008aa}.
  The kinematic coverage was extended to the higher momentum region and to the smaller $\theta$ region from NKS.

\subsection{Three-Track Analysis}
  \label{sec:3_track_analysis}

  We improved a three-track-event analysis with an invariant mass cut and a missing mass cut.
  Events that had two positively charged particles and one negatively charged particle were selected as a candidate for $K^{+} \Lambda$ events.
  $\pi^{-}$ and $p$ are chosen by mass square cuts.
  The invariant mass of $p\pi^{-}$ pair was computed to select $\Lambda$.
  To check consistency of kinematics, the missing mass of X was evaluated and neutron candidate was selected, where X indicates the rest particle in the reaction of $\gamma + d \rightarrow K^{+} + \Lambda + X$.

\begin{figure}[htbp]
    \begin{center}
      \includegraphics[bb=0 0 518 420,width=7.5cm]{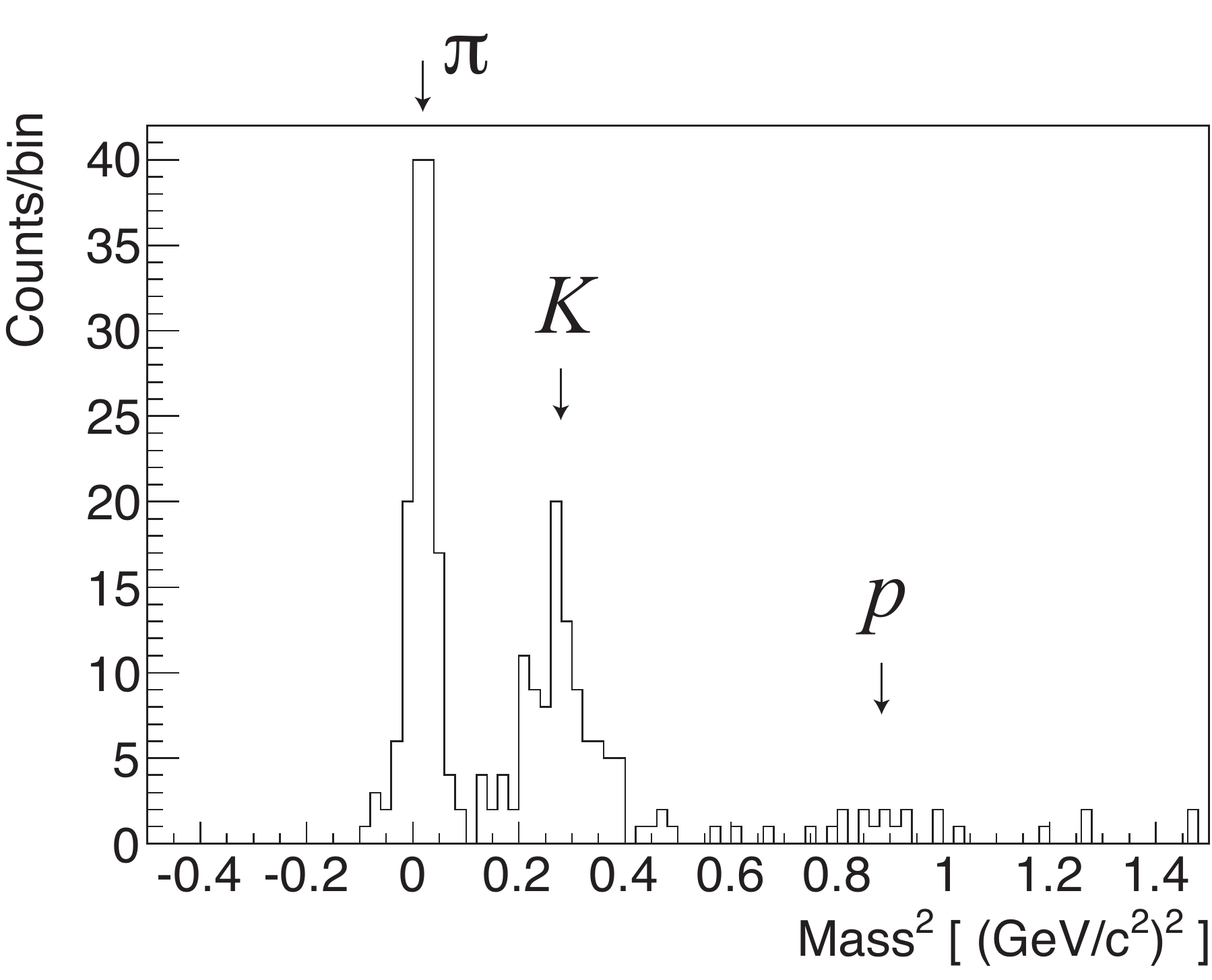}
      \caption{
		Mass square distribution of events corresponding to two positively charged particles and one negatively charged particle, 
		after selecting the invariant mass for $p\pi^{-}$ pair and the missing mass cut of $\gamma + d \rightarrow K^{+} + \Lambda + X$ events (see the text about cut conditions).
		The clear two peaks could be considered as $\pi^{+}$ from $K^{0}_{S}$ decay or $K^{+}$.
		% A bump close to proton mass could be considered as $p$ from $\Delta^{0}$ decay of double delta events (see detail in the text).
		A bump close to proton mass is attributed to $p$ from a background process such as $\gamma + d \rightarrow \pi^{-} + p + p + X$ ($\pi^{-} p$ does not come from $\Lambda$ decay).
              }
      \label{fig:mass_square_from_3track_event}
    \end{center}
  \end{figure}

  Figure~\ref{fig:mass_square_from_3track_event} shows the mass square distribution of positively charged particles.
  A clear peak is seen at the mass square of the $K^{+}$ mass.
  It was clearly separated from $\pi^{+}$.
%  A bump close to proton mass is a contamination of $\gamma + d \rightarrow \pi^{-} + p + p + X$ events.
%  Significant contribution of non-quasi-free (NQF) process, in which two nucleons in the deuteron participate, was observed in our study of the double pion photoproduction~\cite{Hirose:2009aa,Yun-Cheng:2010aa,Kanda:2012aa,Kimura:2013vua,Yamamoto:2014gwa}.
%  In this case, $X$ can be $\pi^{0}$ in NQF process.
%  That could be considered as double delta events ($\gamma + d \rightarrow \Delta^{0} + \Delta^{+}$), as a candidate of such events.
%  The $\Lambda$ candidates of $p \pi^{-}$ pairs have the background from $\Delta^{0} \rightarrow p \pi^{-}$ decay.
%  The rest of particles is considered as the proton from $ \Delta \rightarrow p \pi^{0}$ decay.
  A bump close to proton mass is attributed to $p$ from a background process such as $\gamma + d \rightarrow \pi^{-} + p + p + X$ ($p \pi^{-}$ does not come from $\Lambda$ decay).

\begin{figure}[htbp]
    \begin{center}
      \includegraphics[bb=0 0 523 381,width=7.5cm]{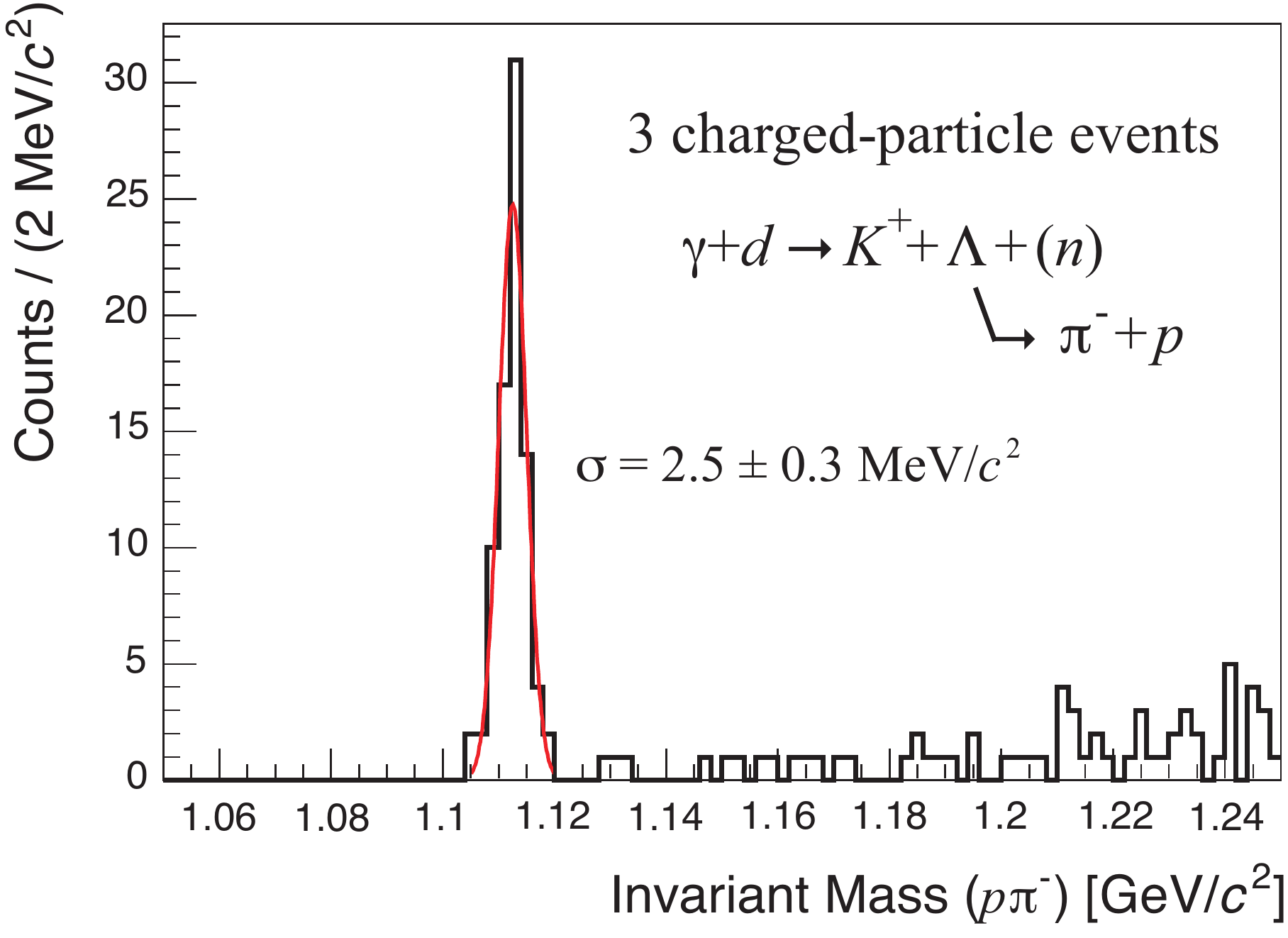}
      \caption{
                      Invariant mass distributions from $p \pi^{-}$ pairs obtained for three-track events of $\pi^{-}, K^{+}$, and $p$. 
                      The fit to the peak with the Gaussian distribution is also shown.
                      The mass resolutions obtained by the Gaussian fit is 2.5 $\pm$ 0.3 MeV/$c^{2}$.
                   }
      \label{fig:inv_mass_three_track}
    \end{center}
  \end{figure}

  Figure~\ref{fig:inv_mass_three_track} is an example of the invariant mass distribution of $p$ and $\pi^{-}$ pairs.
  We chose three-track events (two positively and one negatively charged tracks) and identified particle species with mass square.
  The background of Fig.~\ref{fig:inv_mass_three_track} was less than it in Fig.~\ref{fig:inv_mass_ppi},
  because of less combinatorial background in the three-track event analysis than the two-track analysis.
  The invariant mass resolution was $2.5 \pm 0.3$ MeV/$c^{2}$.

  The invariant mass resolution was estimated by the GEANT4 simulation by varying the position resolution of the drift chambers.
  The results ($\sigma$ of the Gaussian fit) were 1.9, 2.2, and 2.7 MeV/$c^2$ for 0.5, 1.0, and 1.5 times of the position resolution estimated in Section~\ref{sec:position_resolution}, respectively.
  The experimental data ($\sigma = 2.5 \pm 0.3$ MeV/$c^{2}$) were reasonably reproduced by the simulation with 1.0 to 1.5 times of the position resolution ($\sigma = 2.2$ to $2.5$ MeV/$c^{2}$).

  The resolution obtained from Fig.~\ref{fig:inv_mass_ppi} was larger than that from Fig.~\ref{fig:mass_square_from_3track_event} in Section~\ref{sec:3_track_analysis}.
  It is because we used only CDC information in the analysis for Fig.~\ref{fig:inv_mass_ppi} (also for the inclusive $\Lambda$ analysis in Ref.~\cite{Beckford:2016aa}), but all drift chamber information was used in the analysis for Fig.~\ref{fig:inv_mass_three_track} (also for the inclusive $K^{0}_{S}$ analysis and the exclusive analysis).
 The difference in the invariant mass resolutions originated from the different trajectory path lengths.

\subsection{Four-Track Analysis}

  We reported the inclusive measurement of $K^{0}$ or $\Lambda$ in the $\gamma + d$ reaction~\cite{Futatsukawa:2012aa, Beckford:2016aa}.
  The exclusive measurement of $K^{0} \Lambda$ required more statistics to obtain the cross-section.
  We would like to demonstrate the particle identification capability of the $K^{0} \Lambda$ event in this subsection.

\begin{figure}[htbp]
    \begin{center}
      \includegraphics[bb=0 0 225 214,width=7.5cm]{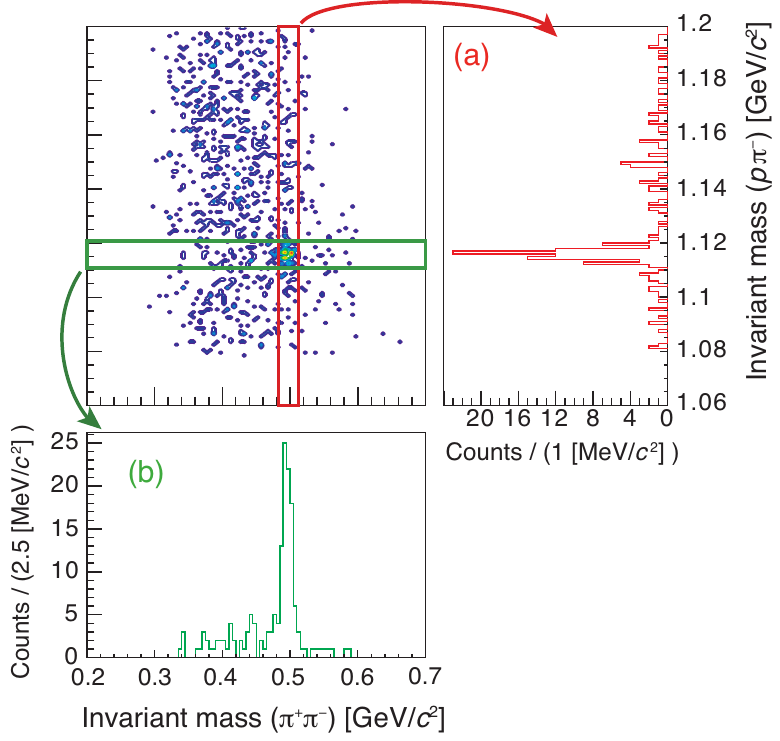}
      \caption{
				Scatter plot of $p \pi^{-}$ and $\pi^{+} \pi^{-}$ Invariant mass ($m_{\pi^{+} \pi^{-}}$ and $m_{p \pi^{-}}$).
				(a) is a $\Lambda$ candidate with a selection of $| m_{K^0} -  m_{\pi^{+} \pi^{-}}| <$ 0.013 GeV/$c^{2}$. 
				(b) is a $K^{0}$ candidate with $| m_{\Lambda} - m_{p \pi^{-}} | <$ 0.005 GeV/$c^{2}$.
				$m_{K^0}$ and $m_{\Lambda}$ represent the $K^{0}$ and $\Lambda$ mass of the PDG values (0.497611 and 1.115683 GeV/$c^2$), respectively. 
				After each selection, the invariant mass distributions were obtained by projecting to each axis.
                   }
      \label{fig:inv_mass_four_track}
    \end{center}
  \end{figure}

  Figure~\ref{fig:inv_mass_four_track} shows the scatter plot of the $p \pi^{-}$ and $\pi^{+} \pi^{-}$ invariant mass and projections to both axes after the mass selection.
  The data are the same set with Ref.~\cite{Futatsukawa:2012aa}.

  Cuts of the decay vertex finding and opening angle selection were applied to the data, similar to the inclusive measurements.
  All combinations in $p \pi^{+} \pi^{-} \pi^{-}$ events were filled in the scatter plot after those cuts.
  The number of $K^{0}\Lambda$ events is 90 in the overlap region of two boxes in Fig.~\ref{fig:inv_mass_four_track}.

%%___________________________________________________________________________________________________________________

\section{Summary}

  A new spectrometer was constructed at the tagged photon beamline in LNS, Tohoku University, to investigate photoproduction of hadrons with liquid deuterium and hydrogen targets.
  The spectrometer was constructed in 2005--2006 and named as NKS2.
  NKS2 consisted of two drift chambers (VDC and CDC), plastic scintillator hodoscopes (IH and OH), EV counters, and a cryogenic target system.
  The acceptance of NKS2 covered the forward region, and it was suitable to measure the photoproduction of $K^{0}$ and $\Lambda$. 
  The DAQ system performed with an efficiency of 70~\% for a 2 kHz trigger rate.

  Strangeness photoproduction data were taken with the liquid deuterium target from 2006 to 2007.
  We upgraded the inner detectors in 2007--2008 and commissioned them in 2009.
  We gathered data in 2010 using the liquid hydrogen and liquid deuterium targets. 
  From the analysis of those data sets, we reported $K^{0}$ and $\Lambda$ photoproduction measurements in the $\gamma$+$d$ reaction.
  NKS2 showed the expected performance.

%%___________________________________________________________________________________________________________________

\section*{Acknowledgments}

  We are grateful to the staff members of the LNS-Tohoku accelerator group.
  We appreciate Prof. T. Takahashi (KEK) for his crucial contribution to the NKS2 experiment in the early stage.
  We would like to express the appreciation to Prof. H. Shimizu (LNS, Tohoku Univ.) for his strong support of the experiment.
  We would like to thank the engineering/technical staff, Mr. N. Chiga (Dept. of Phys., Tohoku Univ.) and Mr. K. Matsuda (LNS, Tohoku Univ.) for their technical support in the construction and development of the spectrometer.
  We thank former graduate students: Mr. J. Fujibayashi, Mr. Y. Miyagi, Mr. N. Maruyama, Mr. N. Terada, Mr. T. Kawasaki, Mr. S. Kiyokawa, Mr. A. Okuyama, and Mr. A. Iguchi (Dept. of Phys., Tohoku Univ.).
  They greatly contributed to the construction of the spectrometer, data taking, and data analysis.
  Additionally, our deepest appreciation goes to Prof. Y. Arai (KEK) and Engineering Department of AMSC Co., Ltd.
  We could not fix the problem of AMT-VME without their support.

  This work is partly supported by JSPS Grant-in-Aid for Creative Research Program (16GS0201), JSPS Core-to-Core program (21002), JSPS Grant-in-Aid for Scientific Research (C) (23540334), and JSPS Grant-in-Aid for Scientific Research (A) (17H01121).
  The pursuit of a problem with the large event size of AMT-VME and the development of the firmware are supported by Young Researcher Support Fund of Faculty of Science, Tohoku University in 2009.

%%___________________________________________________________________________________________________________________

\appendix

\section{Full Runge--Kutta formula for a three-dimensional trajectory}
  \label{appendix:runge_kutta}

  Here, we define two functions: $f_{y}(y', z', B_x, B_y, B_z, p)$ and $g_{z}(y', z', B_x, B_y, B_z, p)$ for the Runge--Kutta formula.
  \begin{eqnarray}
    f_{y}(y', z', B_x, B_y, B_z, p) & \equiv & y'', \\
    g_{z}(y', z', B_x, B_y, B_z, p) & \equiv & z'',
  \end{eqnarray}
  where $y' = \frac{dy}{dx}, z' = \frac{dz}{dx},  y'' = \frac{d^{2}y}{dx^{2}}$, and $z'' = \frac{d^{2}z}{dx^{2}}$.
  The functions $f_{y}$ and $g_{z}$ are described in Ref.~\cite{Wind:1974zz}.
  In both equations, we need to know the value of magnetic field as a function of position.
  We assumed a constant momentum of a traveling charged particle in the formula.

  The Runge--Kutta formula estimates the next position and gradient of the curve from those of the current point.
  The simultaneous differential equation of $x$ is characterized by a step size $\varDelta{x}$.
  We denote the current point by a suffix $i$ and the next point by $i+1$.
  
  The trajectory is computed step by step using the initial values of the starting point $(x, y, z)$, absolute momentum ($p$), and gradients of the curve at the point ($y' = \frac{dy}{dx}, z' = \frac{dz}{dx} $).
  First, the gradient of the curve is estimated by the following equations.
  \begin{equation*}
    \left.
      \begin{aligned}
        y'_{i+1}& = y'_{i} + \frac{1}{6} ( k_1 + 2 k_2 + 2 k_3 + k_4 ) {\varDelta}x , \\
        z'_{i+1}& = z'_{i} + \frac{1}{6} ( j_1 + 2 j_2 + 2 j_3 + j_4 ) {\varDelta}x , \\
         x_{i+1}& =  x_{i} +           \Delta x , \\
         y_{i+1}& =  y_{i} + y'_{i} \, \Delta x  + \frac{1}{6} ( k_1 + k_2 + k_3 ) (\Delta x)^2 , \\
         z_{i+1}& =  z_{i} + z'_{i} \, \Delta x  + \frac{1}{6} ( j_1 + j_2 + j_3 ) (\Delta x)^2 , 
      \end{aligned}
    \right\}
  \end{equation*}
where ${\varDelta}x$ is the step size for the calculation of the $x$ axis and $k_n$ and $j_n (n=1, 2, 3, 4)$ are computed by the recurrence equations.

$k_{1}$ and $j_{1}$ are given by
  \begin{equation*}
    \left.
      \begin{aligned}
        k_{1}& = f_y(y'_{i}, z'_{i}, B_{1x}, B_{1y}, B_{1z}, p ) , \\
        j_{1}& = g_z(y'_{i}, z'_{i}, B_{1x}, B_{1y}, B_{1z}, p ) , 
      \end{aligned}
    \right\}
  \end{equation*}
  where $(B_{1x}, B_{1y}, B_{1z})$ is the component of $\bm{B}$ at $(x_i, y_i, z_i)$.

  $k_{2}$ and $j_{2}$ are given by
  \begin{equation*}
    \left.
      \begin{aligned}
        k_{2}& = f_y( y'_{i}, z'_{i}, B_{2x}, B_{2y}, B_{2z}, p ) , \\
        j_{2}& = g_z( y'_{i}, z'_{i}, B_{2x}, B_{2y}, B_{2z}, p ) , 
      \end{aligned}
    \right\}
  \end{equation*}
  where $(B_{2x}, B_{2y}, B_{2z})$ is the component of $\bm{B}$ at
  \begin{equation*}
    \left.
      \begin{aligned}
         x& = x_{i} + \frac{1}{2}           \varDelta x                                     , \\
         y& = y_{i} + \frac{1}{2} y'_{i} \, \varDelta x + \frac{1}{8} k_1 (\varDelta x)^{2} , \\
         z& = z_{i} + \frac{1}{2} z'_{i} \, \varDelta x + \frac{1}{8} j_1 (\varDelta x)^{2} .
      \end{aligned}
    \right\}
  \end{equation*}

  $k_{3}$ and $j_{3}$ are given by
  \begin{equation*}
    \left.
      \begin{aligned}
        k_{3}& = f_y( y'_{i}, z'_{i}, B_{3x}, B_{3y}, B_{3z}, p ) , \\
        j_{3}& = g_z( y'_{i}, z'_{i}, B_{3x}, B_{3y}, B_{3z}, p ) , 
      \end{aligned}
    \right\}
  \end{equation*}
  where $(B_{3x}, B_{3y}, B_{3z})$ is the component of $\bm{B}$ at
  \begin{equation*}
    \left.
      \begin{aligned}
         x& = x_{i} + \frac{1}{2}           \varDelta x                                     , \\
         y& = y_{i} + \frac{1}{2} y'_{i} \, \varDelta x + \frac{1}{4} k_2 (\varDelta x)^{2} , \\
         z& = z_{i} + \frac{1}{2} z'_{i} \, \varDelta x + \frac{1}{4} j_2 (\varDelta x)^{2} .
      \end{aligned}
    \right\}
  \end{equation*}

  $k_{4}$ and $j_{4}$ are given by
  \begin{equation*}
    \left.
      \begin{aligned}
        k_{4}& = f_y( y'_{i}, z'_{i}, B_{4x}, B_{4y}, B_{4z}, p ) , \\
        j_{4}& = g_z( y'_{i}, z'_{i}, B_{4x}, B_{4y}, B_{4z}, p ) , 
      \end{aligned}
    \right\}
  \end{equation*}
  where $(B_{4x}, B_{4y}, B_{4z})$ is the component of $\bm{B}$ at
  \begin{equation*}
    \left.
      \begin{aligned}
         x& = x_{i} +           \varDelta x                                     , \\
         y& = y_{i} + y'_{i} \, \varDelta x + \frac{1}{2} k_3 (\varDelta x)^{2} , \\
         z& = z_{i} + z'_{i} \, \varDelta x + \frac{1}{2} j_3 (\varDelta x)^{2} .
      \end{aligned}
    \right\}
  \end{equation*}

It is better to adopt a parameter $\lambda (\equiv 1/p)$ in the functions $f_y$ and $g_z$ during the fitting process to avoid the problem of `divided by zero' on a computer.

%%___________________________________________________________________________________________________________________

\section*{References}

\bibliography{NKS2_bibfile}

\end{document}